\documentclass[english,pre,a4paper,aps,reprint]{revtex4-2}
\usepackage[T1]{fontenc}
\usepackage{textcomp}
\usepackage{amsmath}
\usepackage{amssymb}
\usepackage{graphicx}

\makeatletter
\usepackage{hyperref}

\makeatother

\usepackage{babel}
\begin{document}
\title{Stochastic quantum trajectories demonstrate the Quantum Zeno Effect
in open spin 1/2, spin 1 and spin 3/2 systems}
\author{Sophia M. Walls, Julien M. Schachter, Haocheng Qian and Ian J. Ford}
\address{Department of Physics and Astronomy, University College London, Gower
Street, London, WC1E 6BT, United Kingdom}
\email{Correspondence to: sophia.walls.17@ucl.ac.uk}

\begin{abstract}
We investigate the Quantum Zeno Effect in spin 1/2, spin 1 and spin
3/2 open quantum systems undergoing Rabi oscillations,revealing unexplored
features for the spin 1 and spin 3/2 systems. The systems interact
with an environment designed to perform continuous measurements of
an observable, driving the systems stochastically towards one of the
eigenstates of the corresponding operator. The system-environment
coupling constant represents the strength of the measurement. Stochastic
quantum trajectories are generated by unravelling a Markovian Lindblad
master equation using the quantum state diffusion formalism. These
are regarded as a more appropriate representation of system behaviour
than consideration of the averaged evolution since the latter can
mask the effect of measurement. Complete positivity is maintained
and thus the trajectories can be considered as physically meaningful.
The Quantum Zeno Effect is investigated over a range of measurement
strengths. Increasing the strength leads to greater system dwell in
the vicinity of the eigenstates of the measured observable and lengthens
the time taken by the system to return to that eigenstate, thus the
Quantum Zeno Effect emerges. For very strong measurement, the Rabi
oscillations resemble randomly occurring near-instantaneous jumps
between eigenstates. The trajectories followed by the quantum system
are heavily dependent on the measurement strength which other than
slowing down and adding noise to the Rabi oscillations, changes the
paths taken in spin phase space from a circular precession into elaborate
figures-of-eight. For spin 1 and spin 3/2 systems, the measurement
strength determines which eigenstates are explored and the Quantum
Zeno Effect is stronger when the system dwells in the vicinity of
certain eigenstates compared to others.
\end{abstract}
\maketitle

\section{Introduction}

The Quantum Zeno Effect (QZE), also known as the Turing Paradox, is
the suppression of the unitary time evolution of a system brought
about by repeated measurements. The first general derivation of the
QZE was performed by Degasperis \emph{et al.} in 1974 \citep{degasperis1974}
but the effect was formally characterised by Misra and Sudarshan \citep{misra1977a},
who showed in 1977 that the probability that a system should remain
in an initial state will approach unity as the frequency of projective
measurements made on it tends to infinity. The QZE was named after
the Greek philosopher Zeno's paradox, which suggested that if an arrow
were continuously observed, it would (arguably) appear to be motionless
and thus would never reach its target. Any disturbance in the rate
of change of a quantum system as a result of measurement can be considered
to be a demonstration of the QZE or the Quantum Anti-Zeno Effect (QAZE),
the latter corresponding to the opposite of the QZE and manifesting
itself as the speeding up of the system dynamics rather than the slowing
down. It has also been demonstrated that the QZE may be exhibited
through a series of frequent, partial, incomplete measurements, such
that an incomplete wavefunction collapse leads the system to be more
likely to remain in its initial state \citep{zhang2019,peres1990a}.

A notable motivation for understanding the QZE is that it might be
used to stabilise quantum systems in a particular state such that
properties of the system such as entanglement and coherence might
be maintained \citep{oliveira2022,salih2021}. The stability of such
properties is of great importance in quantum error correction algorithms
or quantum teleportation, as well as in many other quantum computational
fields \citep{chen2020,li2018,lin2022,long2022}. A recent use of
the QZE was demonstrated by Blumenthal \emph{et al.} \citep{blumenthal2021}.
It was shown that the joint evolution of at least two non-interacting
qubits could be achieved through the continuous measurement of one
qubit, such that the operation becomes a multi-qubit entangling gate.
A similar effect was found by Nodurft \emph{et al.} whereby polarisation
entanglement could be generated between two initially unentangled
photons in coupled waveguides through the QZE \citep{nodurft2022}.
Finally, the QZE has been frequently studied in electron spins in
quantum dots, trapped ions and nuclear spins, and Markovian and non-Markovian
open quantum system dynamics \citep{leppenen2022,patsch2020,bethke2020,madzik2020}.

The conceptual difficulty with studying the QZE theoretically is that
the conventional quantum mechanical framework consists of two distinct
regimes. The unitary time evolution of the quantum system governed
by the internal Hamiltonian is described by the deterministic, time-reversal
symmetric Schr\"{o}dinger equation, whilst measurement of the quantum
system is introduced as an interruption of the unitary evolution of
the quantum system in a discontinuous, indeterministic and irreversible
manner via the Born rule \citep{norsen2017}. With the consequences
of measurement of the system being absent from the dynamical equation
of motion and average quantities masking the effect of measurement,
it is not straightforward to characterise how the unitary evolution
is influenced by measurement \citep{norsen2017,jacobs2014a}.

To reveal the effects of measurement on a system's unitary evolution,
an approach that enables the study of single, physically realistic
quantum trajectories is required. Many experimental and theoretical
works have investigated quantum trajectories \citep{murch2013,jordan2013,bellini2022,gross2022,minev2019}.
Presilla \textit{et al} compared the experimental results of the QZE
found by Itano \textit{et al} on a two-level system with the theoretical
predictions formulated through the path-integral and QSD formalisms
\citep{presilla1996,itano1990}. It was found that that the experimental
results of Itano \textit{et al} correspond to a high measurement coupling
strength between the system and the measuring apparatus. Snizhko \emph{et
al.} recently illustrated, through the study of quantum trajectories,
that the QZE includes a number of transition stages leading to an
increase in qubit survival probability \citep{snizhko2020}. Gambetta
\emph{et al.} also studied the QZE of a superconducting charged qubit
coupled to a transmission line resonator with a Rabi control drive,
undergoing weak homodyne measurements, which illustrated the competition
between the measurement drive and the Rabi oscillation system dynamics
\citep{gambetta2008}.

Approaches exist where individual trajectories are employed to generate
the evolution of the (reduced) density matrix of a system averaged
over environmental conditions, but often these are not regarded as
physically realistic: for example the Stochastic Liouville-von Neumann
equation is based on an evolving ensemble of density matrices that
do not preserve unit trace and thus are not physically meaningful
\citep{devega2017}. Quantum jump trajectories can also be generated
which consist of intervals of piecewise deterministic evolution interrupted
by stochastic, discontinuous jumps \citep{christie2022,gardiner1992}.
More recently, a new method of generating trajectories, namely \textsl{jump-time
unravellings }has emerged whereby quantum trajectories are bundled
together and averaged over at specific points in time where jumps
occur \citep{gneiting2021}. In this work, we instead turn to Quantum
State Diffusion (QSD), also known as weak measurement \citep{percival1998}.
QSD offers continuous stochastic quantum trajectories and a dynamical
treatment of measurement instead of instantaneous wavefunction collapse.
Bauer \textit{et al} and Spiller have demonstrated that the QZE is
a phenomenon that naturally arises in a spin-1/2 system treated within
the QSD framework \citep{bauer2015a} \citep{spiller1994}. We extend
this by demonstrating the QZE, and its dependence on the strength
of measurement coupling, in spin systems higher than spin 1/2 and
reveal novel features that emerge only in the higher spin systems.

QSD considers quantum systems that are in contact with an environment
that causes the system reduced density matrix to \textit{diffuse}
across its parameter space. Environmental effects in QSD can drive
phenomena such as decoherence and the (continuous) collapse of the
system to a particular eigenstate, all within a single dynamical framework.
An average over all possible diffusive trajectories then yields the
evolution of the density matrix of the quantum system as given by
the Lindblad equation, at least for Markovian dynamics.

For our purposes, we consider the environment to represent a measuring
apparatus, and the strength of the interaction between the quantum
system and environment to be a measurement coupling that brings about
the diffusion of the system towards a stochastically selected eigenstate.
Rather than the system collapsing discontinuously to an eigenstate,
as is the case with projective measurements, the measurement process
occurs in a continuous and non-instantaneous manner. An approximation
to instantaneous wavefunction collapse is obtained in the limit of
infinitely strong system-environment coupling. Near-instantaneous
quantum jumps are recovered in this limit (not to be confused with
the intrinsically discontinuous Poissonian jump unravelling of a quantum
master equation) \citep{gambetta2008,christie2022,gardiner1992}.
The QSD evolution equation for the system therefore contains not only
the unitary time evolution brought about, for example, by a system
Hamiltonian, but also the effect of measurement.

The probabilistic nature of the measurement process is represented
as stochastic noise in the QSD framework, generating dynamics described
by stochastic differential equations (SDEs) or It\^{o} processes when
the stochasticity is Markovian \citep{percival1998}. The origin of
such a stochasticity might be interpreted as pseudo-random rather
than truly indeterministic since the noise that is introduced into
the system dynamics could be a reflection of the underspecified state
of the environment with which the system is interacting, rather like
the noise experienced by an open system like a Brownian particle in
classical mechanics. Within such a framework we consider the stochastically
evolving density matrix to be a physically real property of the quantum
system \citep{penrose2014a,ghirardi1986a,bohm1952b,saunders2010,everett1957,bassi2013,pusey2012,lombardi2021a},
instead of merely a tool to calculate the average evolution or representing
the (subjective) state of our knowledge \citep{petersen1963,fuchs2014,pitowsky2005,healey2012,harrigan2010a,kochen1967,brukner2002,mermin2013,rovelli2021},
which is still the subject of debate, also in the context of quantum
trajectories \citep{maudlin2019,leifer2014,norsen2017,hiley2000,wiseman1996,gambetta2003}.
Nevertheless, results obtained from QSD are consistent with the Born
rule, and are therefore compatible with the axioms of quantum mechanics.

In this study of the QZE, we employ QSD to study spin 1/2, spin 1
and spin 3/2 systems, each undergoing Rabi oscillations while coupled
to a measuring device that monitors the component of spin along the
$z$-axis. We derive stochastic differential equations for the expectation
values of the three Cartesian components of the spin operator, but
reinterpret these quantities as stochastic properties of the system
associated with individual quantum trajectories as opposed to averages
over multiple realisations of the noise or equivalently over an ensemble
of adoptable density matrices. The quantum trajectories of each of
these systems are used to observe the effects of increased measurement
coupling on the unitary dynamics of the system. In particular, the
stochastic quantum trajectories are analysed to calculate the probabilities
of system residence in the vicinity of $z$-spin eigenstates, the
average time taken by the system to return to such an eigenstate,
and how these quantities depend on the measurement strength.

The plan for the paper is as follows. Section \ref{sec:Stochastic_Diff_Eqns}
summarises the key ideas of QSD and SDEs, and Section \ref{sec:System_set-up}
provides a specification of the parametrisation and dynamics of the
spin systems. Section \ref{sec:Results} describes the quantum trajectories
obtained using such a framework and through an analysis of such trajectories
presents insights into the QZE. Our conclusions are given in Section
\ref{sec:Discussion_Conclusion}.

\section{Quantum State Diffusion and a Stochastic Lindblad Equation \label{sec:Stochastic_Diff_Eqns}}

\subsection{Open Quantum Systems and the Kraus Operator Formalism \label{sec:kraus_operators}}

Open quantum systems are ubiquitous since most systems can be found
interacting with their environment. The complexity of this interaction
as well as uncertainty in the initial state of the environment mean
that an element of randomness enters into the description of the environment's
influence on a quantum system. This motivates a description of the
open quantum system using a randomly evolving (reduced) density matrix. 

In order to develop these ideas, consider the commonly used evolution
of the density matrix $\rho(t)$ defined through the action of the
super-operator $S$ in a time interval $dt$. Such a mapping of states
of the quantum system must preserve the unit trace and positivity
for the quantum state to be considered physically meaningful \citep{breuer2007}.
The action of the super-operator on $\rho$ can be expressed as

\begin{align}
S[\rho(t)]= & \rho(t+dt)=\sum_{j}M_{j}(dt)\rho(t)M_{j}^{\dagger}(dt).\label{eq:Kraus_operators-1}
\end{align}
The Kraus operators $M_{j}(dt)$ represent all possible transitions
brought about by the system dynamics together with system-environment
interactions and can in principle be derived from terms in the Hamiltonian.
In order to preserve the trace of $\rho$ they must satisfy the completeness
relation $\sum_{j}M_{j}^{\dagger}M_{j}=\mathbb{I}$. The Kraus operators
depend on $dt$ and, for continuous evolution of the density matrix,
must differ by a small perturbation from the identity operator $\mathbb{I}$.
For simplicity we use the set of Kraus operators $M_{l}\equiv M_{k\pm}(dt)=\frac{1}{\sqrt{2}}(\mathbb{\mathbb{I}}+A_{k\pm})$,
where $k$ labels a particular Lindblad channel $k$, with the $\pm$
subscript representing two possible measurement outcomes or transitions
associated with the interaction. We employ $A_{k\pm}=-iH_{s}dt-\frac{1}{2}L_{k}^{\dagger}L_{k}dt\pm L_{k}\sqrt{dt}$
where $H_{s}$ is the system Hamiltonian and $L_{k}$ the operator
associated with the $k$th Lindblad channel \citep{matos2022,clarke2022,gross2018,jacobs2014a}.
The use of two Kraus operators per Lindblad channel is reminiscent
of the approach used by Wiseman, namely Kraus operators $\Omega_{k1}=\sqrt{dt}L_{k}$
and $\Omega_{k0}=\mathbb{\mathbb{I}}-(iH_{s}+\frac{1}{2}L_{k}^{\dagger}L_{k})dt$
\citep{wiseman1996}, such that $\Omega_{k1}=\frac{1}{\sqrt{2}}\left(M_{k+}-M_{k-}\right)$
and $\Omega_{k0}=\frac{1}{\sqrt{2}}\left(M_{k+}+M_{k-}\right)$. Note
that the Kraus operators $\Omega_{k0,1}$ and $M_{k\pm}$ yield the
same Lindblad master equation but differ in their stochastic realisations
of the trajectories, where the former produces discontinuous, 'jump-like'
trajectories and the latter diffusive, continuous trajectories.

Such a positive map then results in the ubiquitous Lindblad master
equation describing the Markovian evolution of the reduced density
matrix $\rho$ averaged over the measurement outcomes:

\begin{equation}
d\rho=-i[H_{s},\rho]dt+\sum_{k}\left(L_{k}\rho L_{k}^{\dagger}-\frac{1}{2}\{L_{k}^{\dagger}L_{k},\rho\}\right)dt.\label{eq:Lindblad_eqn-1}
\end{equation}
 The density operator $\rho$ may be regarded as representing a statistical
ensemble of the pure states that could be adopted by an open quantum
system after a measurement has been performed. As such, the Lindblad
master equation describes the average evolution of this ensemble under
the action of the Kraus operators.

Another possible interpretation of Eq. (\ref{eq:Kraus_operators-1}),
when written as 
\begin{align}
\rho(t+dt)= & \sum_{l}p_{l}(dt)\frac{M_{l}(dt)\rho(t)M_{l}^{\dagger}(dt)}{Tr(M_{l}(dt)\rho(t)M_{l}^{\dagger}(dt))},\label{eq:kraus_alt}
\end{align}
is that the dynamics may be represented as stochastic transitions
from $\rho(t)$ to $M_{l}(dt)\rho(t)M_{l}^{\dagger}(dt)/Tr(M_{l}(dt)\rho(t)M_{l}^{\dagger}(dt))$
that occur with a probability $p_{l}(dt)=Tr(M_{l}(dt)\rho(t)M_{l}^{\dagger}(dt))$
that depends on the density matrix of the system at time $t$. Eq.
\ref{eq:Kraus_operators-1} represents the average evolution of the
density matrix as a result of the action of all the Kraus operators
on $\rho(t)$. Stochastic quantum trajectories might then be generated
through a series of transitions employing a sequence of Kraus operators.
This is a strategy that has been used in studies of evolving quantum
systems \citep{wiseman1996,breuer2007}.

Our strategy is to generate trajectories reflecting a single possible
evolution path of $\rho$ rather than an averaged one, and so we take
the stochastic interpretation just mentioned. We are modelling measurement
as an interaction between a system and its environment that affects
the system in a manner not dissimilar to the effect of a host medium
on a Brownian particle. Such an approach can make sense since the
state of the environment is unknown, giving rise to an evolving uncertainty
in the state of the quantum system. To reflect such an uncertainty
it would be natural to consider an ensemble of density matrices at
each point in time $t$, reflecting the multiple possible states of
the system as a result of lack of knowledge of the state of the environment.
Similarly to Matos \textit{et al} \citep{matos2022}, the average
of such an ensemble of density operators may be denoted by the over-bar,
$\bar{\rho}$. The evolution of the ensemble averaged density matrix
over a time increment $dt$ would then be expressed as \citep{matos2022}

\begin{equation}
\bar{\rho}(t+dt)=\sum_{l}M_{l}(dt)\bar{\rho}(t)M_{l}^{\dagger}(dt).\label{eq:ensemble_kraus}
\end{equation}
In contrast, the action of a single Kraus operator on a member $\rho(t)$
of such an ensemble represents one of a set of possible system evolutions:

\begin{equation}
\rho(t+dt)=\rho(t)+d\rho=\frac{M_{l}(dt)\rho(t)M_{l}^{\dagger}(dt)}{Tr(M_{l}(dt)\rho(t)M_{l}^{\dagger}(dt))},\label{eq:single_kraus}
\end{equation}
and a sequence of such transformations can generate a single, possible
evolution path of $\rho$. Note that the preservation of the trace
and positivity of the standard Kraus operator map of Eq. \ref{eq:Kraus_operators-1}
is well-known \citep{breuer2007}. Trace preservation is apparent
also in Eq. \ref{eq:single_kraus}. However, preservation of positivity
in the action of a single Kraus operator of the form $M_{k\pm}(dt)=\frac{1}{\sqrt{2}}(\mathbb{\mathbb{I}}+A_{k\pm})$
on one member of an ensemble of density matrices in Eq. \ref{eq:single_kraus}
remains to be demonstrated and will be shown in the following.

Consider a density matrix $\rho(t)$ at a time $t$, specified to
be completely positive and hence to have only non-negative eigenvalues.
The determinant of such a density matrix is $\det(\rho(t))=\prod_{i}\lambda_{i}$
where $\lambda_{i}$ denote the eigenvalues of $\rho(t)$. The determinant
of $\rho(t)$ is thus positive. Now we consider the determinant of
the density matrix at time $t+dt$, namely $\rho(t+dt)$ generated
according to Eq. (\ref{eq:single_kraus}) by a Kraus operator of the
form stated above $M=C(\mathbb{I}+A)$, where $A$ is infinitesimal
(in the sense that $A\to0$ as $dt\to0$) and $C$ is a constant,
such that $\tilde{M}=M/C=\mathbb{I}+A$ differs infinitesimally from
the identity $\mathbb{I}$. We write
\begin{equation}
\rho(t+dt)=\frac{\tilde{M}\rho(t)\tilde{M}^{\dagger}}{Tr(\tilde{M}\rho(t)\tilde{M}^{\dagger})},\label{eq:reduced map}
\end{equation}
and employ the identity $\det(\exp A)=\exp(TrA)$ such that $\det(\mathbb{I}+A)\approx1+Tr(A)$
for infinitesimal $A$, whereby
\begin{align}
\det(\rho(t+dt)) & =\frac{\det(\tilde{M})\det(\rho(t))\det(\tilde{M}^{\dagger})}{Tr(\tilde{M}\rho(t)\tilde{M}^{\dagger})}\nonumber \\
 & \approx\frac{(1+Tr(A))\det(\rho(t))(1+Tr(A^{\dagger}))}{Tr(\tilde{M}^{\dagger}\tilde{M}\rho(t))}\nonumber \\
 & \approx\frac{(1+Tr(A+A^{\dagger}))\det(\rho(t))}{Tr((\mathbb{I}+A+A^{\dagger})\rho(t))}\nonumber \\
 & =\frac{(1+Tr(A+A^{\dagger}))\det(\rho(t))}{1+Tr((A+A^{\dagger})\rho(t))}\nonumber \\
 & \approx\det(\rho(t))[1+Tr(A+A^{\dagger})\nonumber \\
 & \qquad\quad-Tr((A+A^{\dagger})\rho(t))].\label{eq:kraus_positivity}
\end{align}
Next we note that an unacceptable map would produce a density matrix
with at least one negative eigenvalue. We can therefore identify unphysical
dynamics if a density matrix is generated whose determinant passes
through zero at some point in time. This includes cases where an odd
number of eigenvalues change sign simultaneously (such that the determinant
becomes negative) or where an even number do so (such that the determinant
goes through a cusp at zero).

So let us examine the dynamics represented by Eq. (\ref{eq:kraus_positivity}).
It can be demonstrated that the determinant of $\rho(t+dt)$ can never
go to zero within a finite time-frame, under certain conditions. We
express the change in the determinant of the density matrix $d\det(\rho)=\det(\rho(t+dt))-\det(\rho(t))$
as an It\^{o} process: $d\det(\rho)=\det(\rho)(adt+bdW)$, where $dW$
is a Wiener increment and $a$ and $b$ are specified functions. Letting
$y=\ln(\det(\rho)$), the following SDE for $y$ may be obtained:
\begin{align}
dy= & \frac{1}{\det(\rho)}d\det(\rho)+\frac{1}{2}b^{2}\det(\rho)^{2}\left(-\frac{1}{\det(\rho){}^{2}}\right)dt\nonumber \\
 & =adt+bdW-\frac{1}{2}b^{2}dt.
\end{align}
The unacceptable behaviour $\det(\rho)\to0$ corresponds to $y\to-\infty$
and unless $a$ or $b$ possess singularities it is clear that this
can only emerge, if at all, as $t\to\infty$. Thus the dynamics of
the single Kraus operator map, Eq. (\ref{eq:single_kraus}), with
$M=C(\mathbb{I}+A)$ and employing a non-singular, infinitesimal $A$,
are physically acceptable.

Employing the set of Kraus operators from before, $M_{k\pm}=\frac{1}{\sqrt{2}}(\mathbb{\mathbb{I}}+A_{k\pm})$
and $A_{k\pm}=-iH_{s}dt-\frac{1}{2}L_{k}^{\dagger}L_{k}dt\pm L_{k}\sqrt{dt}$,
\citep{matos2022,clarke2022,breuer2007} yields the following Lindblad
equation:

\begin{equation}
d\bar{\rho}=-i[H_{s},\bar{\rho}]dt+\sum_{k}\left(L_{k}\bar{\rho}L_{k}^{\dagger}-\frac{1}{2}\{L_{k}^{\dagger}L_{k},\bar{\rho}\}\right)dt,\label{eq:Lindblad_eqn}
\end{equation}
which explicitly describes the evolution of an average over the ensemble
of density matrices. We generate stochastic trajectories by \textit{unravelling}
Eq. (\ref{eq:Lindblad_eqn}) in the manner of Eq. \ref{eq:single_kraus}
such that an equation concerning a single member of the ensemble of
density matrices $\rho(t)$ can be formulated. We next derive such
an equation for $\rho(t)$ in the form of an It\^{o} process.

\subsection{Unravelled Stochastic Lindblad Equation \label{sec:Unravelling}}

We unravel the deterministic Lindblad equation by implementing the
stochastic evolution of $\rho$ according to the set of available
transitions 
\begin{equation}
\rho(t+dt)=\frac{M_{k\pm}(dt)\rho(t)M_{k\pm}^{\dagger}(dt)}{Tr(M_{k\pm}(dt)\rho(t)M_{k\pm}^{\dagger}(dt))},\label{eq:single kraus pm}
\end{equation}
each adopted with probability $p_{k\pm}(dt)=Tr(M_{k\pm}(dt)\rho(t)M_{k\pm}^{\dagger}(dt))$.
The outcome is an It\^{o} process for $\rho$ involving independent
Wiener increments $dW_{k}$ (in the case of more than one Lindblad
operator) \citep{matos2022}:

\begin{align}
d\rho & =-i[H_{s},\rho]dt+\sum_{k}\Big((L_{k}\rho L_{k}^{\dagger}-\frac{1}{2}\{L_{k}^{\dagger}L_{k},\rho\})dt\nonumber \\
 & \qquad+\left(\rho L_{k}^{\dagger}+L_{k}\rho-Tr[\rho(L_{k}+L_{k}^{\dagger})]\rho\right)dW_{k}\Big).\label{eq:SME}
\end{align}
Solutions to this equation describe possible stochastic quantum trajectories
of the reduced system density matrix $\rho$, each associated with
a particular realisation of the environmental noises $\{dW_{k}\}$
\citep{helmer2009a}, or equivalently, an initial environmental state.
The ensemble-averaged density matrix $\bar{\rho}$ then corresponds
to taking an average over the noise and hence over all possible trajectories
and, as a result, Eq. (\ref{eq:Lindblad_eqn}) is recovered. If the
system were closed instead of open, interactions between the system
and the environment would vanish and we would be left with the first
term on the right-hand side; namely the von Neumann equation. A similar
stochastic evolution equation for $\rho$ has been derived by Jacobs
by analogy with classical measurement theory \citep{jacobs2014a}.
Under this approach, however, the noise is interpreted as a quantum
fluctuation stemming from the uncertainty in the measurement: a similar
approach is taken by Gambetta \emph{et al} \citep{gambetta2008}.
In contrast, we consider such a noise to reflect the probabilistic
nature of interactions with an underspecified environment causing
the quantum system to evolve in a stochastic manner.

\section{System specification \label{sec:System_set-up}}

\subsection{Stochastic Differential Equation Parameters}

We consider a spin system with a Hamiltonian given by

\begin{equation}
H_{s}=\epsilon S_{x},\label{eq:spin_hamiltonian}
\end{equation}
where $\epsilon$ is a positive constant and the operator $S_{x}$
represents the $x$-component of the spin: for example in the case
of a spin 1/2 system we have $S_{i}=\frac{1}{2}\sigma_{i}$ where
$\sigma_{i}$ denotes a Pauli spin matrix. The spin operators satisfy
the condition $[S_{i},S_{j}]=i\epsilon_{ijk}S_{k}$ where $\epsilon_{ijk}$
is the Levi-Civita symbol. We shall consider cases of spin 1/2, spin
1 and spin 3/2 systems where the matrix representations of the $S_{x}$,
$S_{y}$ and $S_{z}$ operators accordingly have dimensions 2, 3 and
4, respectively. The Hamiltonian in Eq. (\ref{eq:spin_hamiltonian})
produces Rabi oscillations, as we shall see.

\begin{figure}
\begin{centering}
\includegraphics[width=1\columnwidth]{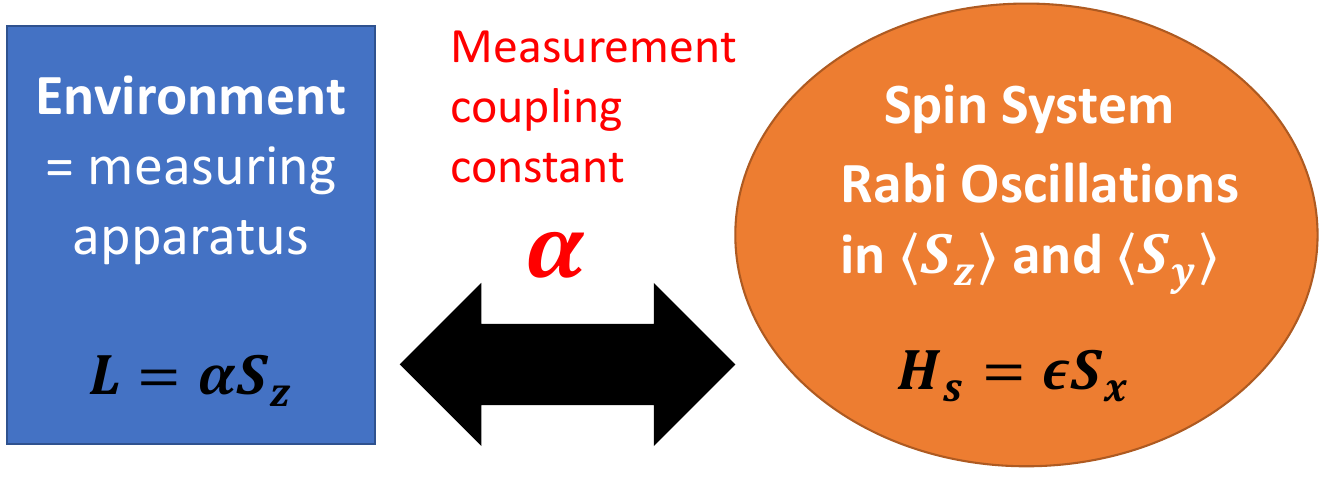}
\par\end{centering}
\caption{An open quantum spin system interacting with an environment with a
strength defined by the coupling constant $\alpha$. The environment
acts as a measurement apparatus, measuring the system's spin along
the $z$-axis. }
\label{fig:sys_env_setup}
\end{figure}

The desired effect of the environment on the system is that it should
act as a measuring apparatus. Diffusive evolution of the system's
$z$-component of spin, to be defined shortly, towards one of the
eigenstates of the $S_{z}$ operator, can be achieved using a single
Lindblad operator in Eq. (\ref{eq:SME}) of the form 

\begin{equation}
L=\alpha S_{z},\label{eq:envir_coupling_operator}
\end{equation}
where the constant $\alpha$ denotes the degree of coupling between
the system and its environment through the operator $S_{z}$ and can
be regarded as the strength of measurement. Figure \ref{fig:sys_env_setup}
illustrates the features of the system and its environment. Utilising
the above expressions for the Lindblad operator and the system Hamiltonian
in Eq. (\ref{eq:SME}), the following SDE is obtained for the evolution
of the reduced density matrix of the system:

\begin{align}
d\rho & =-i\epsilon[S_{x},\rho]dt+\alpha^{2}\left(S_{z}\rho S_{z}-\frac{1}{2}(\rho S_{z}^{2}+S_{z}^{2}\rho)\right)dt\nonumber \\
 & \qquad+\alpha\left(\rho S_{z}+S_{z}\rho-2\langle S_{z}\rangle\rho\right)dW.\label{eq:d_rho}
\end{align}
The single Wiener increment $dW$ is a Gaussian noise with a mean
of zero and a variance $dt$, and $\langle S_{z}\rangle$ is the $z$-component
of spin, defined by $Tr(S_{z}\rho)$. Note that the latter is not
to be confused with the usual quantum expectation value of $S_{z}$
which in our notation is $Tr(S_{z}\bar{\rho})$, where $\bar{\rho}(t)$
is the density matrix at time $t$ averaged over all possible quantum
trajectories. The expectation value represents an average of a (projectively)
measured system property taking into account the quantum mechanical
randomness of measurement outcome as well as the range of quantum
states that might be adopted by an open system when coupled to an
uncertain environment. By extension, $\langle.\rangle=Tr(.\rho)$
could be interpreted as a conditional average projective measurement
outcome given a specific stochastically evolving density matrix. But
it could instead be regarded simply as a property of the density matrix,
and hence of the physical state, and indeed it is spin components
such as $\langle S_{z}\rangle$ that undergo Rabi oscillations.

Terms in Eq. (\ref{eq:d_rho}) that depend on $\alpha$ represent
the effects of measurement on the system, brought about by its interaction
with the environment. Notice that for $\epsilon=0$ the stationary
states for the measurement dynamics are $\rho=\vert m_{z}\rangle\langle m_{z}\vert$
where the $\vert m_{z}\rangle$ are eigenstates of $S_{z}$ satisfying
$S_{z}\vert m_{z}\rangle=m_{z}\vert m_{z}\rangle$. Starting from
an arbitrary initial state, the coupling to the environment captured
by the Lindblad operator in Eq. (\ref{eq:envir_coupling_operator})
evolves the system towards $z$-spin eigenstates as desired.

\subsection{Density Matrix Parametrisation }

The density matrix of the open quantum system is to be parametrised
by a set of variables $\{x_{i}\}$ which evolve stochastically according
to the general It\^{o} processes \citep{ito1944}:

\begin{equation}
dx_{i}=A_{i}dt+\sum_{k}B_{ik}dW_{k},\label{sdes}
\end{equation}
where $dW_{k}$ denote the Wiener increments and the $A_{i}$ and
$B_{ik}$ are designated functions. The number of parameters required
to represent the density matrix depends on the system. Three are required
for a density matrix describing a spin 1/2 system. For a spin 1 system,
eight parameters are needed since this is the number of independent
variables required to specify a $3\times3$ Hermitian matrix with
unit trace. Through a similar reasoning fifteen variables are needed
to parametrise the density matrix for a spin 3/2 system.

\subsubsection{Spin 1/2}

The density matrix of a spin 1/2 system can be expressed using the
Bloch sphere formalism, namely $\rho=\frac{1}{2}\left(\mathbb{I}+\boldsymbol{r}\cdot\boldsymbol{\sigma}\right)$
with coherence (or Bloch) vector $\boldsymbol{r}=(x,y,z)=Tr(\rho\boldsymbol{\sigma})=\langle\boldsymbol{\sigma}\rangle$
\citep{aerts2014}. SDEs in the form of Eq. (\ref{sdes}) can be derived
starting from
\begin{equation}
dr_{i}=Tr(\sigma_{i}d\rho),\label{eq:sdes 1/2}
\end{equation}
with the insertion of Eq. (\ref{eq:d_rho}). For clarity of presentation,
however, we shall employ a one parameter representation for the special
case of a pure state with $Tr\rho^{2}=1$ or $\vert\boldsymbol{r}\vert=1$,
with coherence vector confined to the $(y,z)$ plane, \emph{i.e.}
\begin{equation}
\rho=\frac{1}{2}(\mathbb{I}+\cos\phi\,\sigma_{z}-\sin\phi\,\sigma_{y}),\label{eq:reduced coh vect}
\end{equation}
namely $x=0$, $y=-\sin\phi$ and $z=\cos\phi$. The SDE for $z$
arising from $dz=Tr(\sigma_{z}d\rho)$ and Eq. (\ref{eq:d_rho}) is

\begin{equation}
dz=-\epsilon\sin\phi dt+\alpha\sin^{2}\phi\,dW.\label{eq:sde_z}
\end{equation}
The Rabi angle $\phi$ represents the angle of rotation of the coherence
vector about the $x$ axis. In the absence of environmental coupling,
the action of the system Hamiltonian $\frac{1}{2}\epsilon\sigma_{x}$
is to increase $\phi$ linearly in time. In the presence of such coupling
the evolution of the Rabi angle represents the effect of measurement
on the unitary dynamics of the system. The SDE for $\phi=\cos^{-1}z$
can be found through use of It\^{o}'s lemma \citep{ito1944}:

\begin{equation}
d\phi=\frac{d\phi}{dz}dz+\frac{1}{2}\beta^{2}\frac{d^{2}\phi}{dz^{2}}dt,\label{eq:Ito's_Lemma}
\end{equation}
where $\beta=\alpha\sin^{2}\phi$. Inserting Eq. (\ref{eq:sde_z})
produces

\begin{equation}
d\phi=\left(\epsilon-\frac{1}{4}\alpha^{2}\sin2\phi\right)dt-\alpha\sin\phi\,dW.\label{eq:spin1/2_eqn-motion}
\end{equation}
When $\alpha=0$ the Rabi angle increases at a constant rate, while
for $\alpha\ne0$ the Wiener noise $dW$ disturbs its evolution. The
average rate of change of $\phi$ is given by
\begin{equation}
\frac{d\langle\phi\rangle}{dt}=\epsilon-\frac{1}{4}\alpha^{2}\langle\sin2\phi\rangle,\label{eq:averaged dphi}
\end{equation}
where the brackets again represent an ensemble average. The QZE will
emerge if the term proportional to $\alpha^{2}$ represents an average
retardation of the rotation. In order to resolve this matter we consider
the Fokker-Planck equation for $p(\phi,t)$, the probability distribution
function (PDF) of the Rabi angle $\phi$:
\begin{equation}
\frac{\partial p}{\partial t}=-\frac{\partial\left((\epsilon-\frac{1}{4}\alpha^{2}\sin2\phi)p\right)}{\partial\phi}+\frac{1}{2}\alpha^{2}\frac{\partial^{2}\left(\sin^{2}\phi\,p\right)}{\partial\phi^{2}}.\label{eq:fpe}
\end{equation}
For small $\alpha$, an approximate stationary PDF may be obtained:

\begin{equation}
p_{{\rm st}}(\phi)\propto1+\frac{3\alpha^{2}}{4\epsilon}\sin2\phi,\label{eq:stationary_pdf}
\end{equation}
which shows that the PDF is disturbed from uniformity, in this regime,
by increasing the environmental coupling $\alpha$, or by reducing
$\epsilon$ and hence slowing the rate of Rabi oscillation for an
isolated system. We shall investigate the shape of the stationary
PDF numerically for a range of $\alpha$ in Section \ref{sec:Results}.
For now, let us notice that using this approximation the average needed
in Eq. (\ref{eq:averaged dphi}) is proportional to
\begin{equation}
\int_{0}^{2\pi}d\phi\left(1+\frac{3\alpha^{2}}{4\epsilon}\sin2\phi\right)\sin2\phi\propto\int_{0}^{2\pi}d\phi\sin^{2}2\phi>0,\label{eq:dphi contribution}
\end{equation}
showing that the effect of measurement at small $\alpha$ is indeed
a mean retardation of the Rabi oscillations, consistent with the Quantum
Zeno Effect.

\subsubsection{Spin 1}

The generalised Bloch sphere formalism is used to represent the density
matrix of a spin 1 system. This permits any $3\times3$ density matrix
to be written in terms of the Gell-Mann matrices $\lambda_{i}$ through

\begin{equation}
\rho=\frac{1}{3}(\mathbb{I}+\sqrt{3}\boldsymbol{R}\cdot\boldsymbol{\lambda}),\label{eq:rho_spin_1}
\end{equation}
where $\boldsymbol{R}=(s,m,u,v,k,x,y,z)$ is an eight dimensional
coherence (or Bloch) vector and the Gell-Mann matrices (see Appendix
\ref{sec:Appendix_A}) form the elements of the vector $\boldsymbol{\lambda}=(\lambda_{1},\lambda_{2},\lambda_{3},\lambda_{4},\lambda_{5},\lambda_{6},\lambda_{7},\lambda_{8})$
\citep{lukach1978}. The following density matrix emerges for the
spin 1 system:

\begin{equation}
\rho=\frac{1}{3}\begin{pmatrix}1+\sqrt{3}u+z & -i\sqrt{3}m+\sqrt{3}s & \sqrt{3}v-i\sqrt{3}k\\
i\sqrt{3}m+\sqrt{3}s & 1-\sqrt{3}u+z & \sqrt{3}x-i\sqrt{3}y\\
\sqrt{3}v+i\sqrt{3}k & \sqrt{3}x+i\sqrt{3}y & 1-2z
\end{pmatrix}.\label{eq:spin1_density_matrix}
\end{equation}
Stochastic quantum trajectories can be produced from the equations
of motion for the eight variables parametrising $\rho$. As before,
these are generated from

\begin{equation}
dR_{i}=\frac{\sqrt{3}}{2}Tr(d\rho\lambda_{i}),\label{eq:spin1_eqns_motion}
\end{equation}
using the properties of the Gell-Mann matrices, and are specified
in Appendix \ref{sec:Appendix_B}. The components of the coherence
vector $\boldsymbol{R}$ are denoted $R_{i}$. 

\subsubsection{Spin 3/2}

In order to construct the density matrix of the spin 3/2 system, an
obvious representation would be to utilise SU(4) generators in the
generalised Bloch sphere representation:

\begin{equation}
\rho=\frac{1}{4}(\mathbb{I}+\boldsymbol{s}\cdot\boldsymbol{\Sigma}).\label{eq:rho_spin_3/2}
\end{equation}
The 15 component coherence vector is $\boldsymbol{s}=(v,e,f,g,h,j,k,l,m,n,o,p,q,s,u)$
and the vector of generators $\boldsymbol{\Sigma}$ has components
$\Sigma_{k}$ with $k=1,15$. However, Aerts\emph{ }and Sassoli de
Bianchi discuss how such a representation beyond the SU(2) generators
could potentially allow the system to explore unphysical states \citep{aerts2014}.
Instead, we use Eq. (\ref{eq:rho_spin_3/2}) with generators $\Sigma_{k}$
constructed using the tensor product of two Pauli matrices: $\Sigma_{k}=\sigma_{i}\otimes\sigma_{j}$:
full details are given in Appendix \ref{sec:Appendix_C}. The following
density matrix is then obtained for the spin 3/2 system:\begin{widetext}
\begin{equation}
\rho=\frac{1}{4}\begin{pmatrix}1+f+p+u & -ie+q-is+v & g+k-il-io & h-ij-im-n\\
ie+q+is+v & 1-f+p-u & h+ij-im+n & g-k-il+io\\
g+k+il+io & h-ij+im+n & 1+f-p-u & -ie-q+is+v\\
h+ij+im-n & g-k+il-io & ie-q-is+v & 1-f-p+u
\end{pmatrix}.\label{eq:spin_3/2_density_matrix}
\end{equation}
\end{widetext}As before, stochastic quantum trajectories can be produced
from the equations of motion of the 15 variables parametrising the
density matrix. SDEs for components of the coherence vector $s_{k}$
are obtained through 

\begin{equation}
ds_{k}=Tr(d\rho\Sigma_{k}),\label{eq:spin3/2_eqn_motion}
\end{equation}
and the details can be found in Appendix \ref{sec:Appendix_D}.

\subsection{Spin Component Stochastic Differential Equations}

The stochastic trajectories of the variables parametrising the density
matrices introduced for the spin quantum systems enable us to describe
the system dynamics exactly, but do not offer much direct physical
insight into the QZE. A more useful way of visualising the behaviour
of such systems is to consider the evolution of the time dependent
quantities $\langle S_{x}\rangle$, $\langle S_{y}\rangle$ and $\langle S_{z}\rangle$.
Recall that these could be interpreted as average values under projective
measurements, hence a set of statistics of the dynamics, though we
prefer to regard them as actual physical properties of the current
quantum state. It is these quantities that perform noisy Rabi oscillations
and we shall refer to them simply as spin components. Two methods
exist to determine their evolution.

The first method simply employs $\langle S_{i}\rangle=Tr(S_{i}\rho)$
such that the spin components can be written in terms of the variables
parametrising the density matrices in Eqs. (\ref{eq:reduced coh vect}),
(\ref{eq:spin1_density_matrix}) and (\ref{eq:spin_3/2_density_matrix}).
Alternatively, by considering $d\langle X\rangle=Tr(Xd\rho)$, where
operators $X$ are various functions of the $S_{i}$, the SDEs that
govern the evolution of the spin components can be found using Eq.
(\ref{eq:d_rho}) and solved. We consider these approaches for the
three spin systems in turn.

\subsubsection{Spin 1/2}

For the special case density matrix in Eq. (\ref{eq:reduced coh vect})
the spin components $\langle S_{y}\rangle$ and $\langle S_{z}\rangle$
can be written in terms of the Rabi angle $\phi(t)$ as follows:
\begin{align}
\langle S_{y}\rangle & =-\frac{1}{2}\sin\phi,\qquad\langle S_{z}\rangle=\frac{1}{2}\cos\phi,\label{eq:spin 1/2 spin components}
\end{align}
so the phase space explored by $\langle S_{y}\rangle$ and $\langle S_{z}\rangle$
is a circle of radius $1/2$. We expect a typical stochastic trajectory
to dwell increasingly in the vicinity of the $z$-spin eigenstates
at $\phi=0$ and $\pi$ as the measurement strength increases, and
will demonstrate this in Section \ref{sec:Results}. We also expect
the mean rate of passage between the two eigenstates to reduce as
the measurement strength increases.

The following three SDEs describe the spin components:

\begin{align}
 & d\langle S_{z}\rangle=\epsilon\langle S_{y}\rangle dt+2\alpha\left(\frac{1}{4}-\langle S_{z}\rangle^{2}\right)dW\nonumber \\
 & d\langle S_{x}\rangle=-\frac{\alpha^{2}}{2}\langle S_{x}\rangle dt-2\alpha\langle S_{z}\rangle\langle S_{x}\rangle dW\nonumber \\
 & d\langle S_{y}\rangle=-\epsilon\langle S_{z}\rangle dt-\frac{\alpha^{2}}{2}\langle S_{y}\rangle dt-2\alpha\langle S_{z}\rangle\langle S_{y}\rangle dW.\label{eq:spin_1/2_sde}
\end{align}
For $\alpha=0$ we clearly see the emergence of Rabi oscillations
in $\langle S_{y}\rangle$ and $\langle S_{z}\rangle$. Similarly,
for $\epsilon=0$ we identify stationary states at $\langle S_{x}\rangle=\langle S_{y}\rangle=0$
and $\langle S_{z}\rangle=\pm1/2$, which correspond to the measurement
eigenstates. 

It may be shown that the purity of the system defined by $P=Tr(\rho^{2})$
satisfies
\begin{equation}
dP=\alpha^{2}\left(1-r_{z}^{2}\right)\left(1-P\right)dt+2\alpha r_{z}\left(1-P\right)dW,\label{eq:purity}
\end{equation}
such that purity moves towards $P=1$ and stays there under the given
dynamics. Furthermore, $\langle S_{x}\rangle=0$ is a fixed point
of the dynamics of Eq. (\ref{eq:spin_1/2_sde}) so the special case
considered in Eq. (\ref{eq:reduced coh vect}) describes a more general
asymptotic behaviour. 

\subsubsection{Spin 1}

Similarly, the spin components $\langle S_{x}\rangle$, $\langle S_{y}\rangle$
and $\langle S_{z}\rangle$ can be written in terms of the variables
parametrising the density matrix in Eq. (\ref{eq:spin1_density_matrix}):
\begin{align}
\langle S_{x}\rangle & =\sqrt{\frac{2}{3}}(x+s)\nonumber \\
\langle S_{y}\rangle & =\sqrt{\frac{2}{3}}(m+y)\nonumber \\
\langle S_{z}\rangle & =\frac{u}{\sqrt{3}}+z.\label{eq:spin 1 spin components}
\end{align}
The purity $P$ of the system can be written

\begin{equation}
P=Tr(\rho^{2})=\frac{2}{3}(s^{2}+m^{2}+u^{2}+v^{2}+k^{2}+x^{2}+y^{2}+z^{2})+\frac{1}{3},\label{eq:spin 1 purity}
\end{equation}
and this evolves asymptotically to unity. The dynamics in terms of
spin components and related quantities take the form of eight coupled
stochastic differential equations: \begin{widetext}

\begin{align}
 & d\langle S_{z}\rangle=\epsilon\langle S_{y}\rangle dt+2\alpha\left(\langle S{}_{z}^{2}\rangle-\langle S_{z}\rangle^{2}\right)dW\nonumber \\
 & d\langle S_{x}\rangle=-\frac{\alpha^{2}}{2}\langle S_{x}\rangle dt+\alpha\left(\langle S_{z}S_{x}\rangle+\langle S_{x}S_{z}\rangle-2\langle S_{z}\rangle\langle S_{x}\rangle\right)dW\nonumber \\
 & d\langle S_{y}\rangle=-\epsilon\langle S_{z}\rangle dt-\frac{\alpha^{2}}{2}\langle S_{y}\rangle dt+\alpha\left(\langle S_{z}S_{y}\rangle+\langle S_{y}S_{z}\rangle-2\langle S_{z}\rangle\langle S_{y}\rangle\right)dW\nonumber \\
 & d\langle S{}_{y}^{2}\rangle=-\epsilon\left(\langle S_{y}S_{z}\rangle+\langle S_{z}S_{y}\rangle\right)dt-\alpha^{2}\left(i\langle S_{x}S_{y}S_{z}\rangle+\langle S_{z}^{2}\rangle+\langle S_{y}^{2}\rangle-1\right)dt+\alpha\langle S_{z}\rangle\left(1-2\langle S_{y}^{2}\rangle\right)dW\nonumber \\
 & d\langle S_{z}^{2}\rangle=\epsilon\left(\langle S_{y}S_{z}\rangle+\langle S_{z}S_{y}\rangle\right)dt+2\alpha\langle S_{z}\rangle\left(1-\langle S_{z}^{2}\rangle\right)dW\nonumber \\
 & d\left(\langle S_{y}S_{z}\rangle+\langle S_{z}S_{y}\rangle\right)=2\epsilon\left(\langle S_{y}^{2}\rangle-\langle S_{z}^{2}\rangle\right)dt-\frac{\alpha^{2}}{2}\left(\langle S_{y}S_{z}\rangle+\langle S_{z}S_{y}\rangle\right)dt+\alpha\left(\langle S_{y}\rangle-2\langle S_{z}\rangle\left(\langle S_{y}S_{z}\rangle+\langle S_{z}S_{y}\rangle\right)\right)dW\nonumber \\
 & d\left(\langle S_{z}S_{x}\rangle+\langle S_{x}S_{z}\rangle\right)=-\frac{\alpha^{2}}{2}\left(\langle S_{x}S_{z}\rangle+\langle S_{z}S_{x}\rangle\right)dt+\alpha\left(\langle S_{x}\rangle-2\langle S_{z}\rangle\left(\langle S_{x}S_{z}\rangle+\langle S_{z}S_{x}\rangle\right)\right)dW\nonumber \\
 & d\langle S_{x}S_{y}S_{z}\rangle=i\epsilon\left(\langle S_{y}S_{z}\rangle+\langle S_{z}S_{y}\rangle\right)dt+i\alpha^{2}\left(\langle S_{z}^{2}\rangle+2i\langle S_{x}S_{y}S_{z}\rangle\right)dt+i\alpha\langle S_{z}\rangle\left(1+2i\langle S_{x}S_{y}S_{z}\rangle\right)dW.\label{eq:spin_1_sde}
\end{align}
\end{widetext}Again, the regular Rabi dynamics around a circle in
the phase space spanned by $\langle S_{y}\rangle$ and $\langle S_{z}\rangle$
is apparent when $\alpha=0$. When $\alpha\ne0$ the trajectories
are more complicated. For $\epsilon=0$ the eigenstates of the $z$-spin
are stationary: the relevant spin components being $\langle S_{x}\rangle=\langle S_{y}\rangle=0$
and $\langle S_{z}\rangle=\pm1,0$.

\subsubsection{Spin 3/2}

Once again, the spin components can be written in terms of the variables
parametrising the density matrix for the spin 3/2 system in Eq. (\ref{eq:spin_3/2_density_matrix})
as follows:

\begin{align}
\langle S_{x}\rangle & =\frac{1}{2}(h+n+\sqrt{3}v)\nonumber \\
\langle S_{y}\rangle & =\frac{1}{2}(\sqrt{3}e-j+m)\nonumber \\
\langle S_{z}\rangle & =\frac{1}{2}f+p.\label{eq:spin_3/2_values}
\end{align}
The purity of the system $P$ may be written

\begin{align}
P & =\frac{1}{4}\Big(1+e^{2}+f^{2}+g^{2}+h^{2}+j^{2}+k^{2}+l^{2}\label{eq:spin_3/2_purity}\\
 & \qquad+m^{2}+n^{2}+o^{2}+p^{2}+q^{2}+s^{2}+u^{2}+v^{2}\Big).\nonumber 
\end{align}
For the spin 3/2 system a closed set of fifteen stochastic differential
equations for the spin components and related quantities could not
be found. Thus Eqs. (\ref{eq:spin_3/2_values}) and the SDEs in Appendix
\ref{sec:Appendix_D} are the only way to solve the dynamics for this
spin system. 

\section{Simulations and Results\label{sec:Results}}

\subsection{Spin 1/2}

The equations of motion for the variables parametrising the density
matrices of the spin 1/2, spin 1 and spin 3/2 systems, given by Eqs.
(\ref{eq:spin1/2_eqn-motion}), (\ref{eq:spin1_eqns_motion}) and
(\ref{eq:spin3/2_eqn_motion}), were solved numerically using the
Euler-Maruyama update method \citep{kloeden1992}. From the evolution
of these variables, stochastic quantum trajectories in the phase space
of the $z$-spin component $\langle S_{z}\rangle$ and the $y$-spin
component $\langle S_{y}\rangle$ were produced. For spin 1/2 we also
generated stochastic trajectories of the Rabi angle $\phi$. We chose
the $x$-spin component to be zero initially, where it remains.

\begin{figure}
\begin{centering}
\includegraphics[width=1\columnwidth]{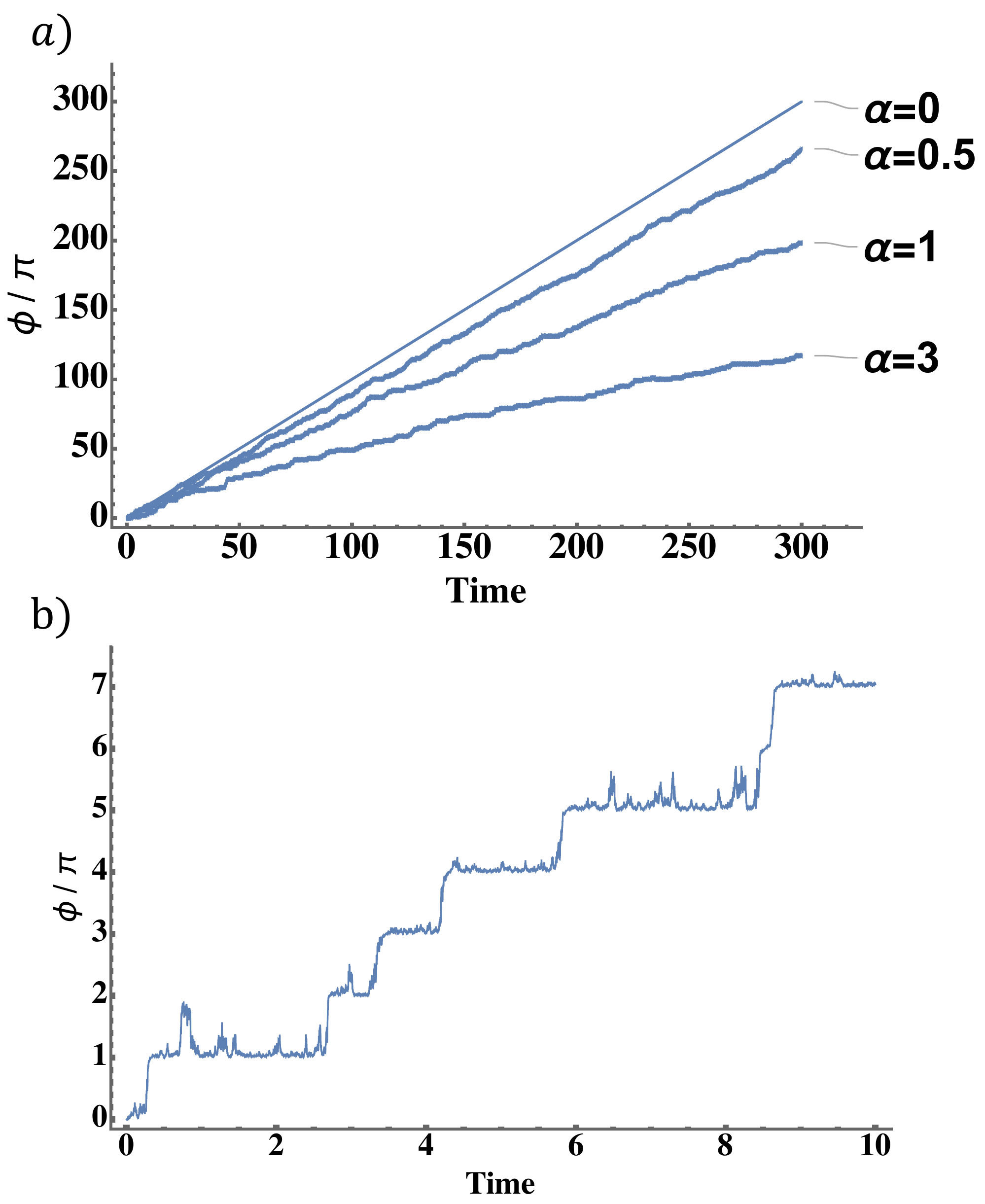}
\par\end{centering}
\caption{a) The evolution of the Rabi angle $\phi$ for a spin 1/2 system at
four values of $\alpha$ with time-step 0.0005, $\epsilon=1$ and
a duration of 300 time units. The mean rate of change is reduced as
the strength of measurement is increased. b) Close-up of the evolution
of $\phi$ for $\alpha=1$ for a duration of 10 time units. Notice
that for this strength of measurement the Rabi angle tends to dwell
in the vicinity of integer multiples of $\pi$, the $z$-spin eigenstates.
\label{fig:rabi_angle}}
\end{figure}

The evolution of the Rabi angle with time for the spin 1/2 system
for various strengths of measurement is shown in Figure \ref{fig:rabi_angle}.
Figure \ref{fig:rabi_angle}a shows that the mean rate of change of
the Rabi angle, namely the mean frequency of Rabi oscillations, decreases
with increasing measurement coupling constant, as was suggested by
Eq. (\ref{eq:averaged dphi}). Such a decrease demonstrates the slowing
down of the unitary dynamics as a result of stronger, or equivalently,
more frequent measurement as is expected for the QZE. Figure \ref{fig:rabi_angle}b
illustrates the dynamics on a finer scale. Behaviour at high measurement
strength resembles to the quantum jumps conventionally considered
to occur between eigenstates of a monitored observable, brought about
by an interaction present in the Hamiltonian.

\begin{figure}
\begin{centering}
\includegraphics[width=1\columnwidth]{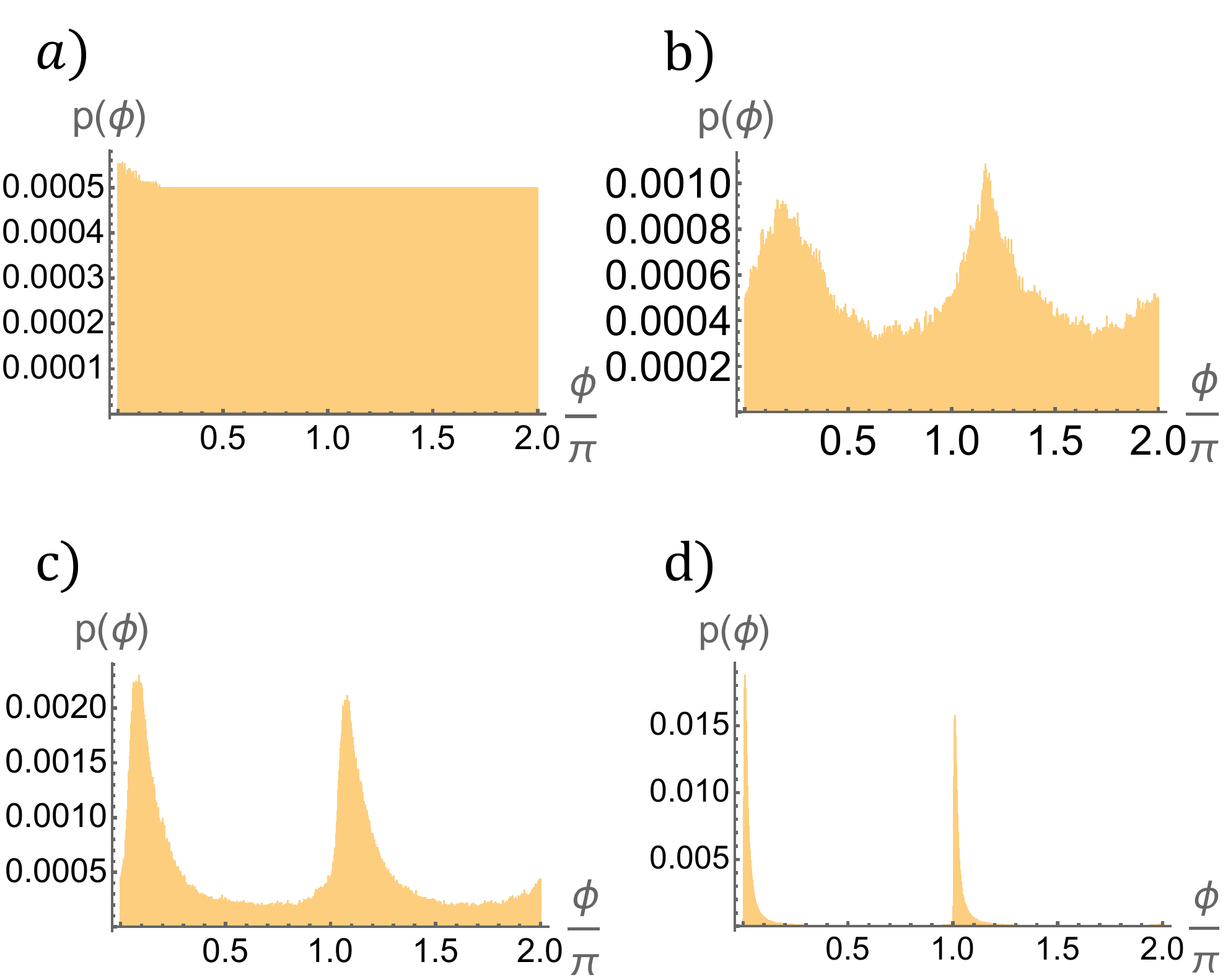}
\par\end{centering}
\caption{\label{fig:pdfs} Stationary probability density functions for the
Rabi angle in the spin 1/2 system, for a range of measurement strengths
$\alpha$, obtained over a duration of 300 time units with a time-step
of 0.0005 and $\epsilon=1$. a) $\alpha=0$, b) $\alpha=0.5$, c)
$\alpha=1$, d) $\alpha=3$. The eigenstates of $S_{z}$ lie at $\phi=0$
and at $\phi=\pi$. Notice the displacement of the peaks above the
eigenvalues, arising from the rotational pull of the system Hamiltonian.}
\end{figure}

Numerically generated stationary PDFs for the Rabi angle are shown
in Figure \ref{fig:pdfs} for an isolated system ($\alpha=0$) and
for three nonzero measurement strengths. Initial states were selected
from a uniform probability density over $\phi$. Angles $\phi=0$
and $\pi$ correspond to the $|\frac{1}{2}\rangle$ and $|\!-\!\frac{1}{2}\rangle$
eigenstates of $S_{z}$, respectively. Values of $\phi$ from the
trajectories are mapped into the range 0 to $2\pi$. It can be seen
that as the measurement strength is increased, the stationary PDFs
become narrower since the measurement dynamics are better able to
localise the system in the vicinity of the eigenstates of $S_{z}$,
resisting the pull of the unitary dynamics induced by the Hamiltonian,
which seek to drive the system into Rabi oscillations. In accordance
with the Born rule, for $\alpha>0$ the stationary PDFs contain two
approximately equivalent peaks around the $|\!-\!\frac{1}{2}\rangle$
and $|\frac{1}{2}\rangle$ eigenstates, since the system should have
an equal probability of localising at either. The PDF in Figure \ref{fig:pdfs}a
is uniform over $\phi$ since the trajectories represent the dynamics
without measurement, namely Rabi oscillations. Note that the stationary
PDF in Figure \ref{fig:pdfs}b is roughly sinusoidal, as suggested
by the PDF for small $\alpha$ given in Eq. (\ref{eq:stationary_pdf}).
Indeed, that approximate result suggested peaks at $\phi=\pi/4$ and
$3\pi/4$: a consequence of the competition between measurement-induced
dwell near the eigenstates and the rotational pull of the Hamiltonian.
As measurement strength increases, the peaks are drawn closer to $\phi=0$
and $\pi$.

\subsection{Spin 1}

\begin{figure}
\begin{centering}
\includegraphics[width=1\columnwidth]{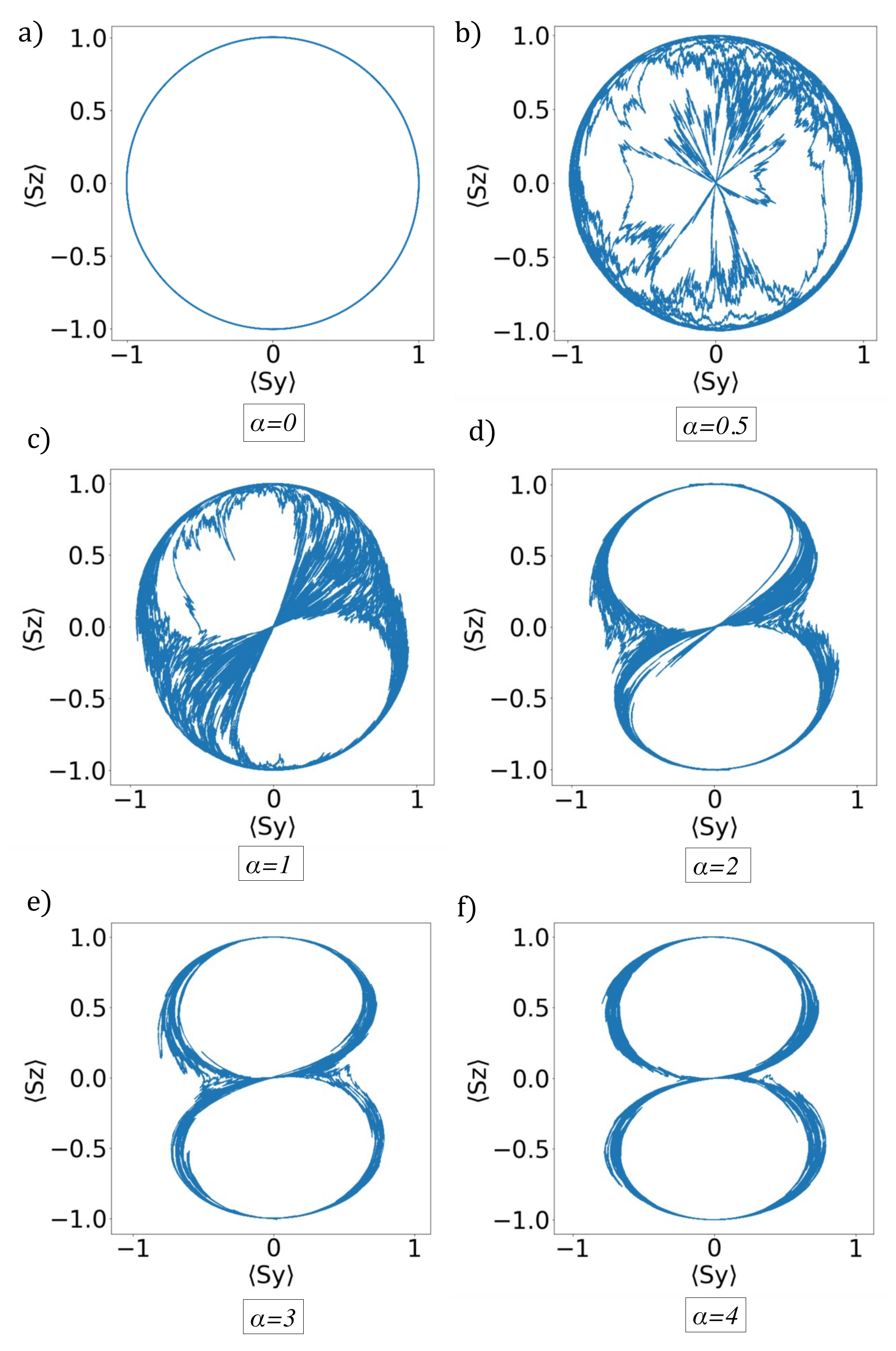}
\par\end{centering}
\centering{}\caption{The path taken through $\langle S_{y}\rangle$, $\langle S_{z}\rangle$
phase space for a spin 1 system with varying measurement strength
$\alpha$, for $\epsilon=1$, time-step 0.0001 and for a duration
of 50 time units. As $\alpha$ increases, the circular path corresponding
to regular Rabi oscillations at $\alpha=0$ becomes disturbed, allowing
visits to the $|0\rangle$ eigenstate at the origin in addition to
the $\vert\!\pm\!1\rangle$ eigenstates at the top and bottom of the
phase space. \label{fig:spin1_trajectories}}
\end{figure}

The evolution of $\langle S_{z}\rangle$ and $\langle S_{y}\rangle$
for a spin 1 system, for different values of the measurement strength
$\alpha$, is illustrated in Figure \ref{fig:spin1_trajectories}.
The system was initialised in the $|\!-\!1\rangle$ spin eigenstate
of the $S_{z}$ operator. Figure \ref{fig:spin1_trajectories}a illustrates
the dynamics when there is no measurement: the Hamiltonian drives
Rabi oscillations in $\langle S_{z}\rangle$ and $\langle S_{y}\rangle$
corresponding to the precession of the spin vector around the orthogonal
$\langle S_{x}\rangle$ axis. The $\vert\!\pm\!1\rangle$ and $\vert0\rangle$
eigenstates of $S_{z}$ lie at $\langle S_{x}\rangle=0$, $\langle S_{y}\rangle=0$
and $\langle S_{z}\rangle=\pm1$ and 0, respectively. If the initial
state of the system had been the $|0\rangle$ eigenstate of $S_{z}$,
then the spin vector would lie on the $\langle S_{x}\rangle$ axis
and would be unable to precess around it, and furthermore, $\langle S_{z}\rangle$
and $\langle S_{y}\rangle$ would be zero.

Disturbance of the circular trajectories can be seen in Figure \ref{fig:spin1_trajectories}b.
As a result of the non-zero measurement strength the system can now
pass near the $|0\rangle$ eigenstate of $S_{z}$, located at the
origin. As the $\alpha$ is increased further in Figure \ref{fig:spin1_trajectories}c-d,
a figure-of-eight pathway forms and the system dwells more often at
the $|0\rangle$ eigenstate until, in Figure \ref{fig:spin1_trajectories}e-f,
visits to this state always occur in between visits to the $\vert\!-\!1\rangle$
and $\vert1\rangle$ eigenstates.

\begin{figure}
\begin{centering}
\includegraphics[width=1\columnwidth]{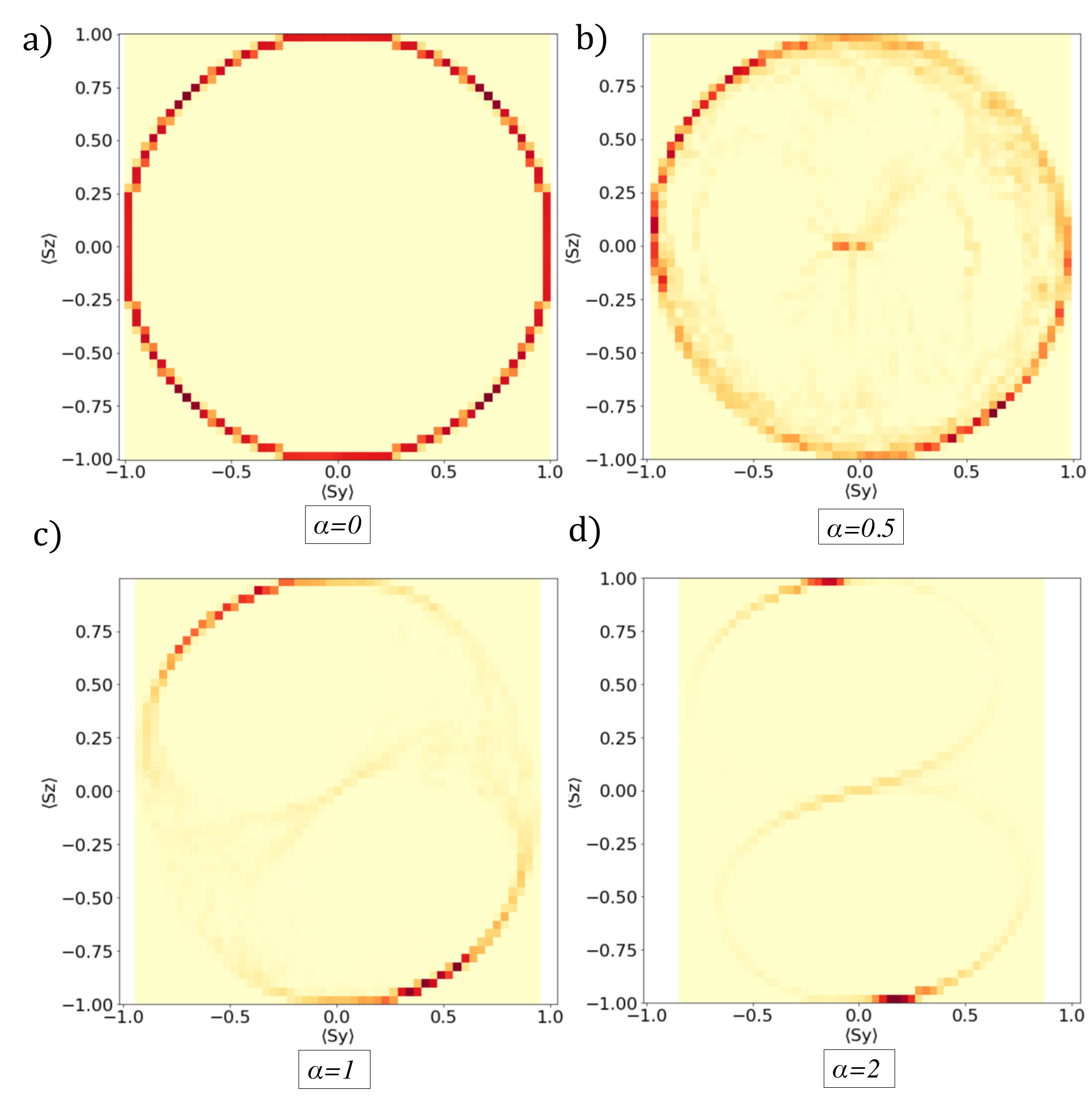}
\par\end{centering}
\begin{centering}
\caption{Density plot (51 by 51 bins) illustrating stationary relative occupation
in the $\langle S_{y}\rangle$, $\langle S_{z}\rangle$ phase space
of the spin 1 system for four values of $\alpha$, obtained from trajectories
with a duration of 100 time units, a time-step of 0.0001 and $\epsilon=1$.
\label{fig:histograms_spin_1}}
\par\end{centering}
\end{figure}

Note that the spin 1 system is not moving at a constant speed around
these figure-of-eight pathways. Instead, the system makes rapid jumps
between the eigenstates, typically in a counter-clockwise direction,
separating periods of dwell in the vicinity of the eigenstates. Movies
of examples of the spin 1 \citep{moviespin1} and spin 3/2 \citep{moviespin32}
system dynamics are available. The behaviour is illustrated in plots
of the relative probability of occupation of patches of the phase
space by the system, shown in Figure \ref{fig:histograms_spin_1}.
The behaviour is similar to that of the spin 1/2 system: a greater
tendency to localise near the eigenstates as $\alpha$ is increased,
while the pull of the Hamiltonian produces a counterclockwise displacement.

\begin{figure}
\begin{centering}
\includegraphics[width=1\columnwidth]{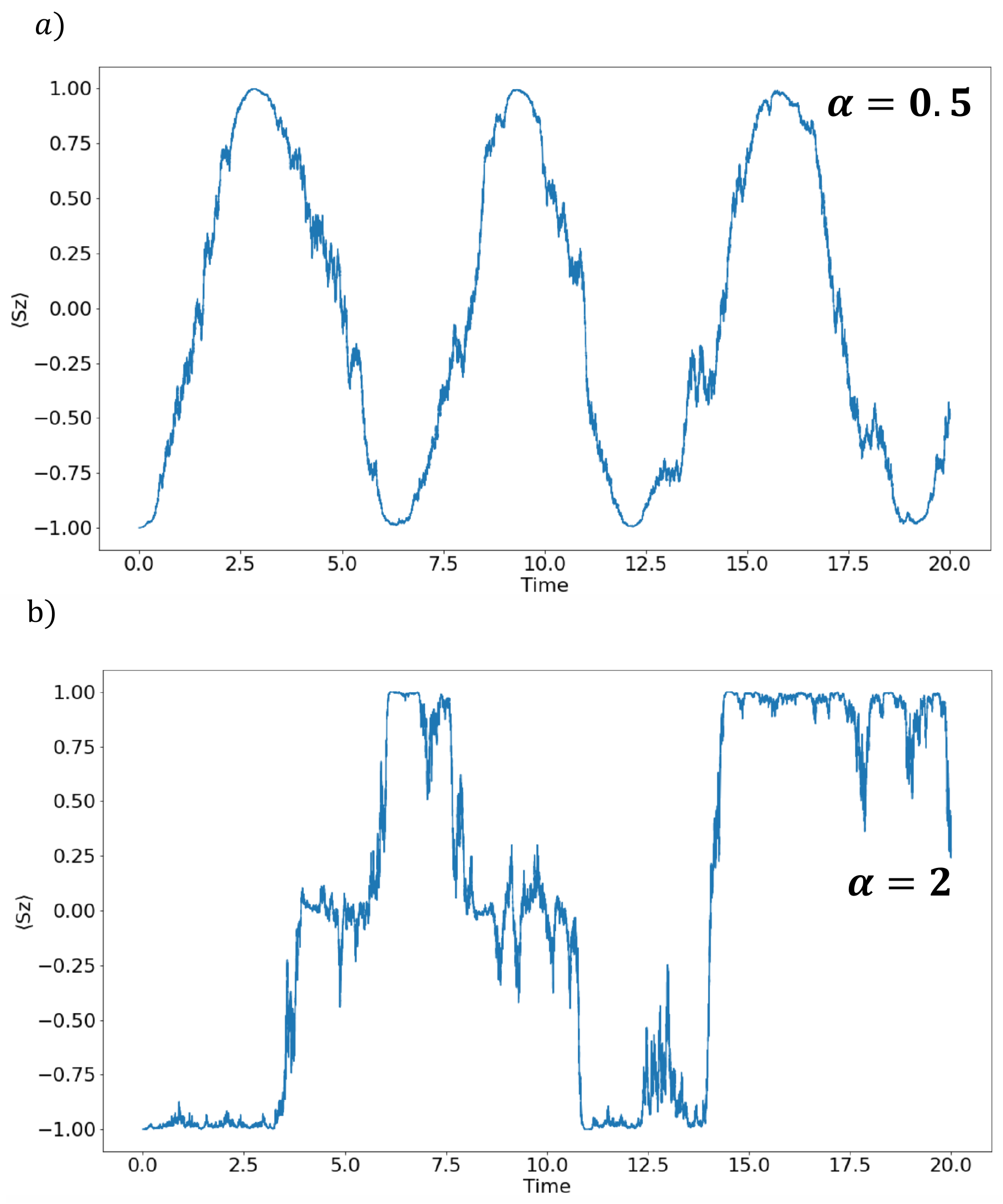}
\par\end{centering}
\caption{Trajectories of $\langle S_{z}\rangle$ for the spin 1 system for
a) $\alpha=0.5$ and b) $\alpha=2$. Slightly noisy Rabi oscillations
at low measurement strength are contrasted with fast jumps between
the eigenstates at $\langle S_{z}\rangle=\pm1$ and 0, with random
waiting times, when the measurement strength is increased.\label{fig:Trajectories-of-Sz}}

\end{figure}

 Figure \ref{fig:Trajectories-of-Sz} further illustrates how increasing
the measurement coupling constant causes the system to dwell for longer
at the three eigenstates of $S_{z}$ and transit more abruptly between
them. The behaviour at high measurement strength increasingly resembles
the textbook behaviour of a system making jumps between eigenstates.

In order to quantify the QZE, we consider the $\alpha$ dependence
of two quantities derived from the trajectories in Figure \ref{fig:spin1_trajectories}.
First, the residence probabilities, which characterise the fraction
of time spent in the vicinity of each eigenstate of $S_{z}$ over
a long simulation. The system is defined to occupy an eigenstate if
the $\langle S_{z}\rangle$ coordinate lies within \textpm 0.1 of
the appropriate eigenvalue. The second quantity considered is the
mean return time. This is the average period from the moment the system
leaves a particular eigenstate (with occupation defined as above)
to the moment it returns to it having visited another eigenstate in
between.

\begin{figure}
\begin{centering}
\includegraphics[width=1\columnwidth]{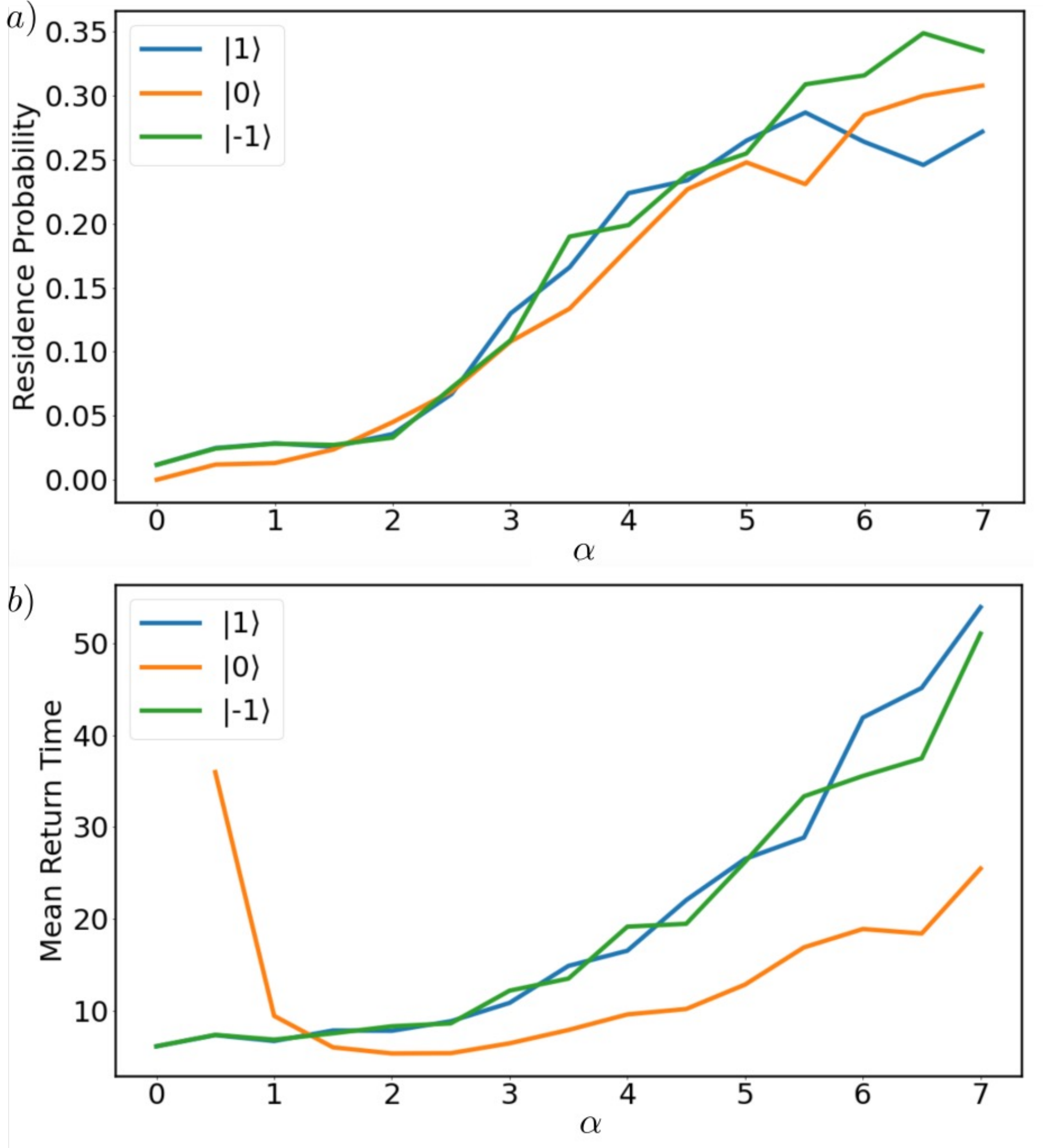}
\par\end{centering}
\caption{Color. a) Residence probabilities, and b) mean return times for the
eigenstates of a spin 1 system, with varying measurement strength
$\alpha$, for $\epsilon=1$, time-step of 0.0001 and for a duration
of 5000 time units. \label{fig:Residence-probabilities-and}}

\end{figure}

Figure \ref{fig:Residence-probabilities-and}a illustrates how residence
probabilities are small for low values of $\alpha$, which is a reflection
of the only slightly disturbed oscillatory behaviour of $\langle S_{z}\rangle$
in Figure \ref{fig:Trajectories-of-Sz}a. But as the coupling strength
is raised, so do the residence probabilities, and this can similarly
be understood by considering the trajectory shown in Figure \ref{fig:Trajectories-of-Sz}b.
Little time is spent in regions far away from the eigenstates. Visits
to the $\langle S_{z}\rangle=0$ eigenstate are less frequent than
to the $\pm1$ eigenstates for small $\alpha$, which is a consequence
of having started the trajectory at $\langle S_{z}\rangle=-1$, but
they occur with approximately equal probability at high values of
$\alpha$.

Figure \ref{fig:Residence-probabilities-and}b shows how the system
takes longer times to return to an eigenstate for stronger measurement.
Again, this is consistent with the behaviour shown in Figure \ref{fig:Trajectories-of-Sz}b.
The mean return time for a given eigenstate is dominated by the period
of dwell at the eigenstate (or eigenstates) to which it moves. The
mean return time to the $\langle S_{z}\rangle=0$ state initially
decreases with $\alpha$ but this is an artefact of the initial condition
for the motion, which makes visits to this eigenstate rare when $\alpha$
is small (as is apparent in Figure \ref{fig:spin1_trajectories}).
The mean return time for the $\langle S_{z}\rangle=0$ eigenstate
is typically less than the return times for $\langle S_{z}\rangle=\pm1$
(roughly half), possibly because direct transitions between the $\langle S_{z}\rangle=1$
and $-1$ states are unlikely. Return to the central eigenstate is
characterised by just one period of dwell at one of the two outer
eigenstates, while return to an outer eigenstate might require waiting
while the system hops between the other two states. The mean return
times in Figure \ref{fig:Trajectories-of-Sz}b can be compared to
the average time between two jumps $\tau$ found from the analytical
expression of the jump rates in Bauer \textit{et al} \citep{bauer2015a}.
Specifically, for $\alpha=7,$ the average time between two jumps
for the $|\pm1\rangle$ eigenstates is $\tau_{|\pm1\rangle}=49$ and
for the $|0\rangle$ eigenstate is $\tau_{|0\rangle}=24.5$, showing
good agreement with our results.

The definition of the vicinity of an eigenstate is, of course, open
to debate\emph{.} The choice we make is simple but is sufficient to
reveal the effects that characterise the QZE. It is natural that the
value of $\langle S_{z}\rangle$ should feature prominently in the
definition, but there are other characteristics of the eigenstates
that could be taken into account. An eigenstate of the $S_{z}$ operator
corresponds to a point in a multidimensional parameter space, and
its vicinity could be defined by putting conditions on a variety of
parameters. We have investigated a more elaborate scheme along these
lines but the resulting residence probabilities and mean return times
are broadly similar to those shown in Figure \ref{fig:Residence-probabilities-and}.
In the interests of simplicity we therefore focus on the value of
$\langle S_{z}\rangle$ alone, and the chosen range of $\pm0.1$ about
eigenvalues.

\subsection{Spin 3/2}

\begin{figure}
\centering{}\includegraphics[width=1\columnwidth]{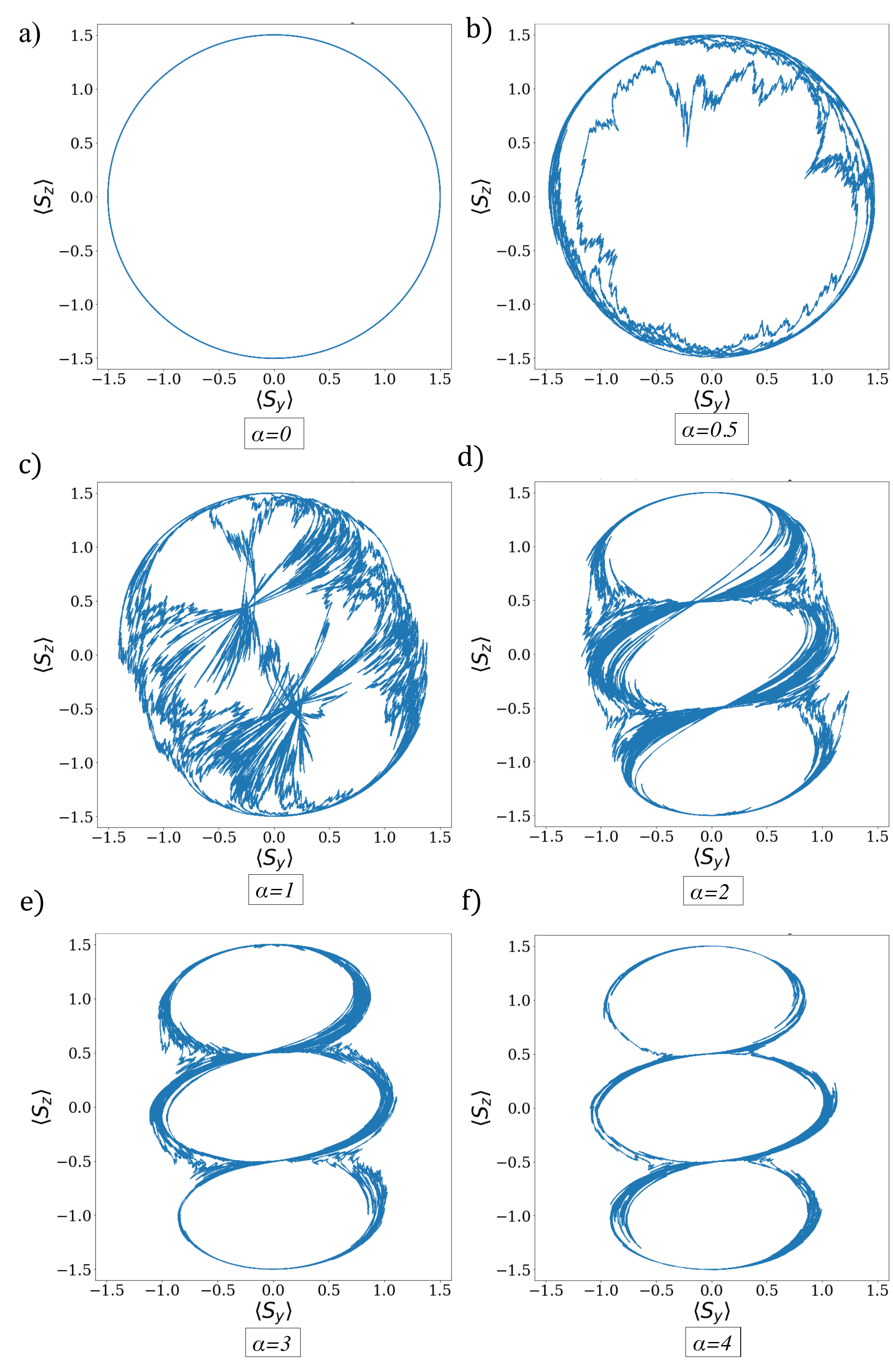}\caption{Paths through the $\langle S_{y}\rangle$, $\langle S_{z}\rangle$
phase space explored by a spin 3/2 system with varying measurement
strength $\alpha$, for $\epsilon=1$, time-step 0.0001 and duration
50 time units. \label{fig:spin_three_half_plots}}
\end{figure}

The effects of measurement on a spin 3/2 system, viewed in terms of
the evolution of spin components $\langle S_{z}\rangle$ and $\langle S_{y}\rangle$,
is illustrated in Figure \ref{fig:spin_three_half_plots} for different
values of the measurement strength $\alpha$. The system was initialised
in the $|\!-\!\frac{3}{2}\rangle$ eigenstate of $S_{z}$. When $\alpha=0$
there is no measurement and, as with the spin 1 system, the Hamiltonian
drives circular trajectories, as depicted in Figure \ref{fig:spin_three_half_plots}a,
passing through the $|\frac{3}{2}\rangle$ and $|\!-\!\frac{3}{2}\rangle$
eigenstates of $S_{z}$ at $\langle S_{y}\rangle=0$ and $\langle S_{z}\rangle=\pm3/2$.
The spin vector precesses around the $\langle S_{x}\rangle$ axis.
As $\alpha$ is increased (Figure \ref{fig:spin_three_half_plots}b-f),
the circular trajectory is disturbed such that the system passes through
further regions of phase space, including the vicinities of the $|\frac{1}{2}\rangle$
and $|\!-\!\frac{1}{2}\rangle$ eigenstates of $S_{z}$ at $\langle S_{y}\rangle=0$
and $\langle S_{z}\rangle=\pm1/2$. When the measurement strength
is sufficiently high, a triple figure-of-eight trajectory emerges,
such that the $|\frac{1}{2}\rangle$ and $|\!-\!\frac{1}{2}\rangle$
eigenstates are encountered between visits to the $|\frac{3}{2}\rangle$
eigenstate and the $|\!-\!\frac{3}{2}\rangle$ eigenstate, and visa-versa.
The system dwells in the vicinities of all four eigenstates of $S_{z}$
for high enough $\alpha$.

\begin{figure}
\begin{centering}
\includegraphics[width=1\columnwidth]{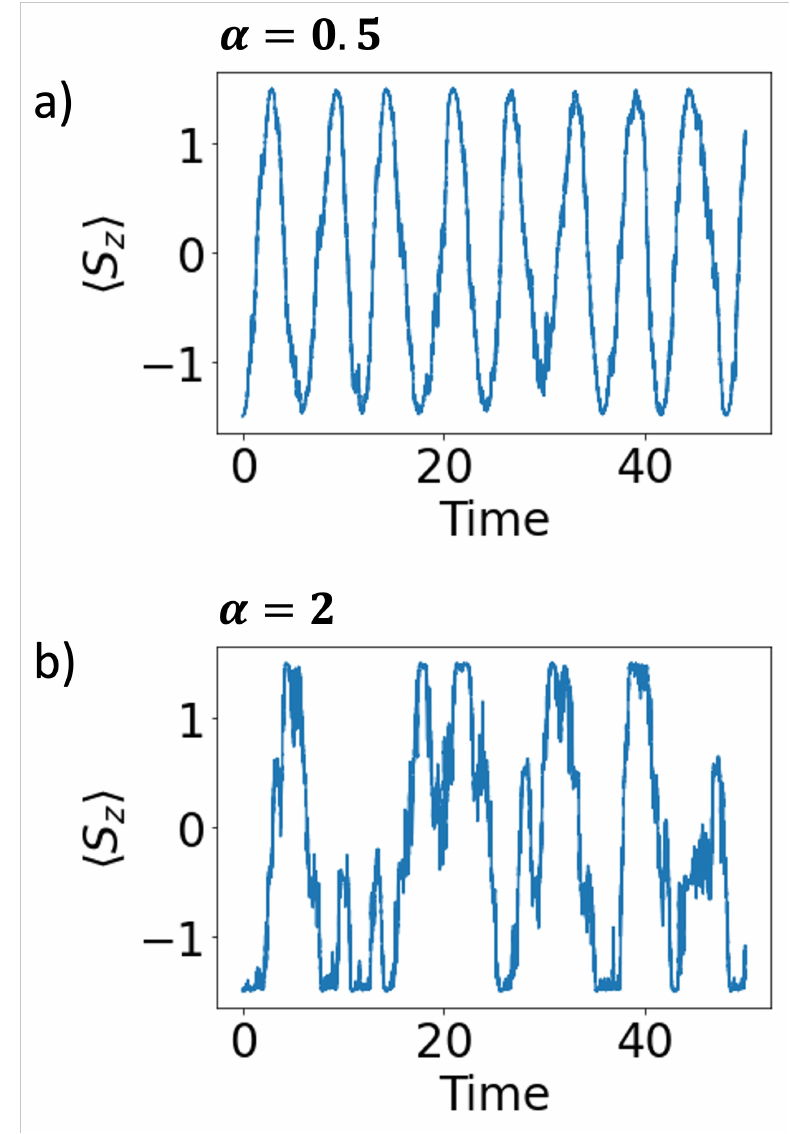}
\par\end{centering}
\caption{Trajectories of $\langle S_{z}\rangle$ for the spin 3/2 system for
a) $\alpha=0.5$ and b) $\alpha=2$. Similarly to the spin 1 system,
noisy Rabi oscillations at low measurement strength are contrasted
with fast jumps between the eigenstates at $\langle S_{z}\rangle=\pm\frac{3}{2}$
and $\langle S_{z}\rangle=\pm\frac{1}{2}$ when the measurement strength
is increased.\label{fig:Trajectories-of-Sz-3half}}
\end{figure}

Similarly to the spin 1 case, Figure \ref{fig:Trajectories-of-Sz-3half}
reveals how increasing $\alpha$ leads to longer dwell at the eigenstates
of $S_{z}$ and more 'jump-like' behaviour in between.

\begin{figure}
\begin{centering}
\includegraphics[width=1\columnwidth]{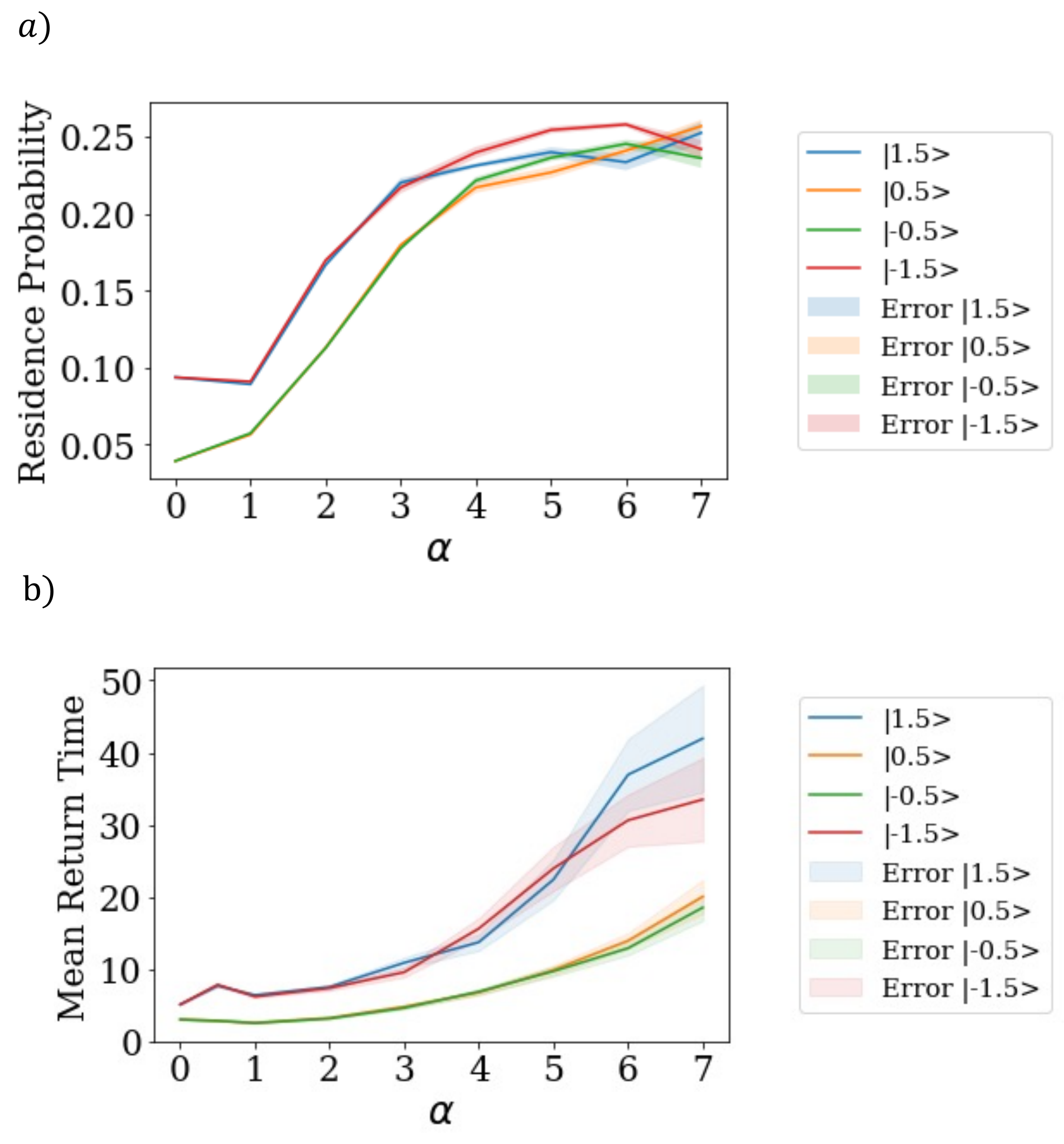}
\par\end{centering}
\caption{Color. The residence probabilities and mean return times for the eigenstates
of a spin 3/2 system, with varying measurement strength $\alpha$,
for $\epsilon=1$, time-step of 0.0001 and for a duration of 5000
time units. \label{fig:return_res_times_three_half}}

\end{figure}

Figure \ref{fig:return_res_times_three_half}a illustrates how the
residence probabilities for the spin 3/2 system depend on $\alpha$.
As with the spin 1 case, the system is defined to occupy an eigenstate
if the $\langle S_{z}\rangle$ coordinate lies within $\pm0.1$ of
the appropriate eigenvalue. At low values of $\alpha$, the Hamiltonian
term dominates and therefore occupation of the $|\frac{1}{2}\rangle$
and $|\!-\!\frac{1}{2}\rangle$ eigenstates of $S_{z}$ is lower than
for the $|\frac{3}{2}\rangle$ and $|\!-\!\frac{3}{2}\rangle$ eigenstates.
For higher values of $\alpha$, the residence probabilities for all
four eigenstates increase and become roughly equal, in concord with
the Born rule, and a triple figure-of-eight trajectory is followed.
For large $\alpha$, the probability of occupying each eigenstate
lies around 0.25, implying that the system is unlikely to occupy a
point in the phase space outside the vicinity of the eigenstates.
Note that increasing the value of $\alpha$ amplifies the noise term
in Eq. (\ref{eq:d_rho}), introducing greater statistical uncertainty
into the simulation.

Figure \ref{fig:return_res_times_three_half}b shows that as the measurement
coupling constant is raised the mean return times increase, hence
demonstrating the QZE. Notably, the mean return times for the $|\frac{1}{2}\rangle$
and $|\!-\!\frac{1}{2}\rangle$ eigenstates of $S_{z}$ are approximately
half those of the $|\frac{3}{2}\rangle$ and $|\!-\!\frac{3}{2}\rangle$
eigenstates. When the system resides at the $|\frac{1}{2}\rangle$
and $|\!-\!\frac{1}{2}\rangle$ eigenstates, and $\alpha$ is high
enough, there are two states to which it can transfer, as opposed
to one when it resides at the $|\frac{3}{2}\rangle$ and $|\!-\!\frac{3}{2}\rangle$
eigenstates. The $|\!\pm\!\frac{1}{2}\rangle$ lie within the ladder
of eigenstates while the $|\!\pm\!\frac{3}{2}\rangle$ are its termini.
Twice the number of paths for a return to $|\frac{1}{2}\rangle$ compared
with $|\frac{3}{2}\rangle$ suggests half the mean return time. The
analytical average time between two jumps $\tau$, found from the
analytical jump rates in Bauer \textit{et al,} also reflect this trend.
Namely, for $\alpha=7,$ the average time between two jumps for the
$|\pm\frac{3}{2}\rangle$eigenstates is $\tau_{|\pm\frac{3}{2}\rangle}\approx32.7$
and for the $|\pm\frac{1}{2}\rangle$ eigenstates is $\tau_{|\pm\frac{1}{2}\rangle}=14$
\citep{bauer2015a}.

We speculate that higher spin systems will continue this pattern of
behaviour. Systems with large measurement strength will follow stochastic
transitions along a multiple figure-of-eight pathway in the phase
space of $\langle S_{z}\rangle$ and $\langle S_{y}\rangle$. The
behaviour will evolve from regular Rabi oscillations towards a situation
where the system (effectively) jumps stochastically between eigenstates
of the monitored observable.

\begin{figure}
\centering{}\includegraphics[width=0.8\columnwidth]{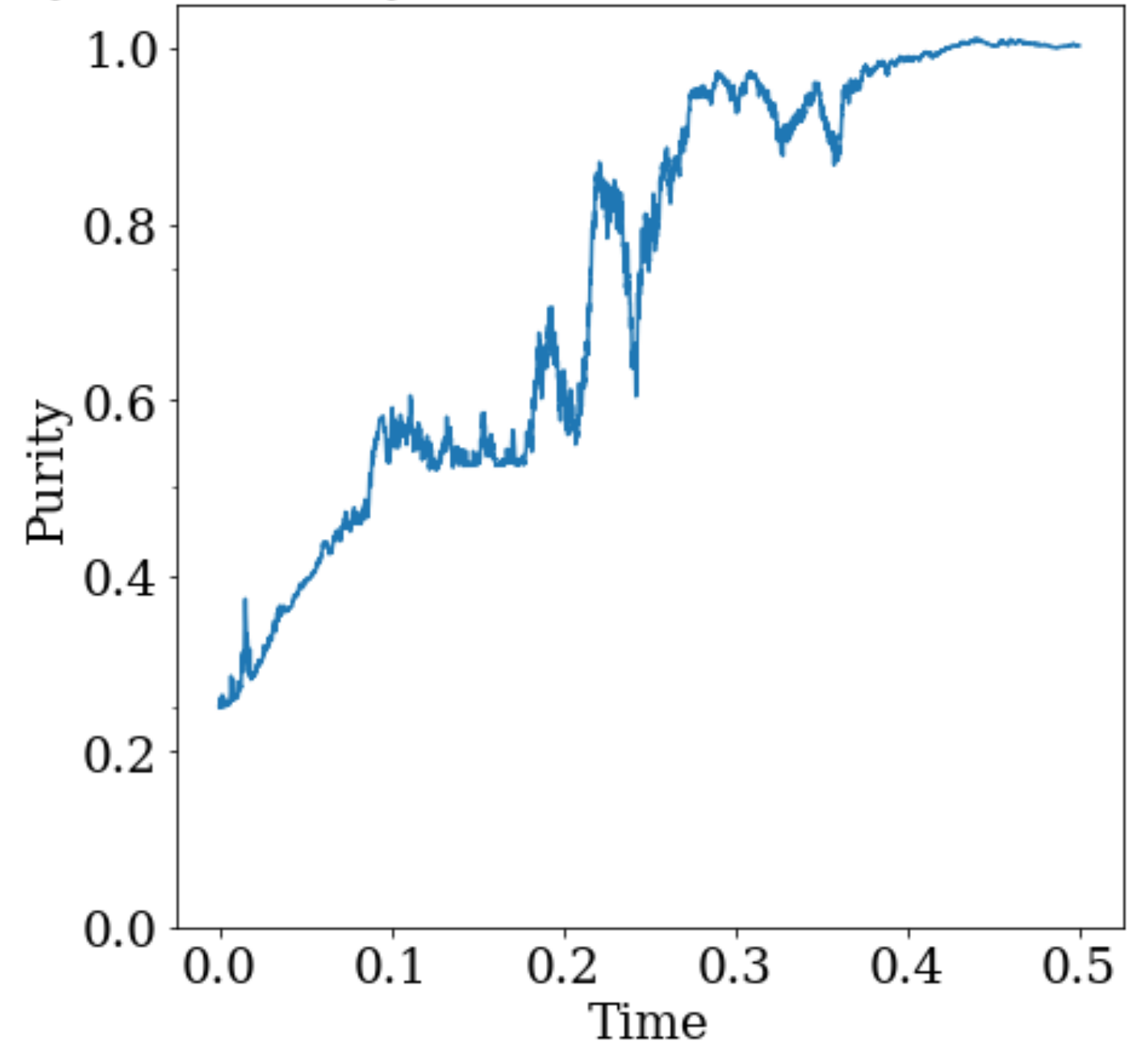}\caption{The evolving purity of a spin 3/2 system for a single quantum trajectory,
starting from a fully mixed state, with $\alpha=3$, $\epsilon=1$,
time-step 0.0001 and for a duration of 0.5 time units. \label{fig:purity_a_3_mixed_three_half}}
\end{figure}

Finally, the evolution of the purity of the spin 3/2 system under
measurement was studied to ensure it remained within the expected
range of $\frac{1}{4}\leq P\leq1$, starting with the system in the
fully mixed state $\rho=\frac{1}{4}\left(|\!-\!\frac{3}{2}\rangle\langle\!-\frac{3}{2}|+|\!-\!\frac{1}{2}\rangle\langle\!-\frac{1}{2}|+|\frac{1}{2}\rangle\langle\frac{1}{2}|+|\frac{3}{2}\rangle\langle\frac{3}{2}|\right)$.
In Figure \ref{fig:purity_a_3_mixed_three_half} it can be seen that
the effect of measurement is to cause the system to purify. The purification
occurs in a continuous but stochastic manner, taking approximately
0.4 time units for the parameters chosen. The approximation of instantaneous
projective measurements and apparent jumps in purity emerges only
in the limit $\alpha\rightarrow\infty$.

It is worth noting that the dynamical framework we use has the effect
that interaction brings about a \emph{disentanglement} of the system
from its environment, namely an increase in purity, which is the opposite
of what is often supposed. However, this is an appropriate outcome
for measurement, where the idea is to convey the system into a (pure)
eigenstate of the appropriate observable. More general system-environment
interactions might change the system purity in different ways, but
to take this into account would require a more explicit representation
of environmental degrees of freedom than that which we have employed.

\section{Discussion and Conclusions\label{sec:Discussion_Conclusion}}

The Quantum Zeno Effect (QZE) has been studied in three open spin
systems undergoing Rabi oscillations generated by Hamiltonian $H_{s}\propto S_{x}$
and coupled to an environment designed to act as a measurement apparatus
for the $S_{z}$ observable. The interaction between the system and
the environment is characterised by a measurement strength $\alpha$.
The QSD approach allows the construction of single, physical, stochastic
quantum trajectories, and can take into account the effects of measurement
in a continuous and explicit manner. The evolution of an open quantum
system is therefore modelled using a stochastically evolving (reduced)
density matrix, $\rho$, preserving unit trace and positivity. The
evolution of the quantum system resembles that of a Brownian particle
diffusing across a phase space, and the density matrix, at each instant,
is interpreted as a physical property of the system. The average over
an ensemble of density matrices represented by the stochastic trajectories
is captured by the noise-averaged Lindblad equation.

Stochastic quantum trajectories for the spin 1/2 system have been
described in terms of the evolution of the Rabi angle. For the spin
1 and spin 3/2 systems, the evolution of the 'expectation values'
$\langle S_{y}\rangle=Tr(\rho S_{y})$ and $\langle S_{z}\rangle=Tr(\rho S_{z})$
were studied. It is important to note that, in spite of the terminology
just used, we regard these spin components as physical properties
rather than statistics of an ensemble and that they characterise a
single stochastic evolution. To demonstrate the QZE, the spin 1 and
spin 3/2 trajectories were analysed to calculate the mean return times
under the dynamics, and the residence probabilities, for the various
eigenstates of $S_{z}$.

The stochastic trajectories for both the spin 1 and spin 3/2 systems
(Figures \ref{fig:spin1_trajectories} and \ref{fig:spin_three_half_plots})
reveal a competition between the unitary dynamics which attempt to
guide the system along a deterministic trajectory in the $\langle S_{y}\rangle$,
$\langle S_{z}\rangle$ phase space, and the non-unitary measurement
dynamics which seek to divert the system stochastically towards the
eigenstates of $S_{z}$, and dwell in their vicinity. For example,
measurement can disturb a deterministic trajectory that passes through
eigenstates with the largest eigenvalues, allowing occasional visits
to the other eigenstates of $S_{z}$, such as the $|0\rangle$ eigenstate
in the case of the spin 1 system and $|\frac{1}{2}\rangle$ and $|\!-\!\frac{1}{2}\rangle$
in the case of the spin 3/2 system. For high enough values of $\alpha$,
the system traverses a figure-of-eight path in the case of the spin
1 system, and a triple-figure-of-eight path for the spin 3/2 system.
We speculate that a similar pattern will be found for higher spin
systems, where for very high values of $\alpha$ the system will simply
appear to jump stochastically between the eigenstates. As such, the
transition pathways available to the quantum system appear to change
according to the strength of the measurement imposed upon it.

The stochastic trajectories generated for all three spin systems for
a range of values of the measurement strength $\alpha$ have clearly
demonstrated the QZE. With increasing $\alpha$, the mean rate of
change of the Rabi angle for the spin 1/2 system decreases (Figure
\ref{fig:rabi_angle}a). Since the Rabi angle describes rotation of
the coherence/Bloch vector about an axis specified by the Hamiltonian,
such a decrease of its mean rate of change demonstrates a slowing
down of the unitary dynamics. Suppression of the unitary dynamics
was also illustrated by the increased mean return time to eigenstates
in the spin 1 and spin 3/2 systems (Figures \ref{fig:Residence-probabilities-and}b
and \ref{fig:return_res_times_three_half}b). Increased measurement
strength causes the system to dwell for longer at each eigenstate,
a clear manifestation of the QZE. The stationary PDFs of the spin
1/2 system (Figure \ref{fig:pdfs}) show that an increase in $\alpha$
narrows the PDFs around the eigenstates of $S_{z}$.

Probabilities of residence near the eigenstates of $S_{z}$ for both
spin 1 and spin 3/2 systems rise to a ceiling with increasing measurement
coupling constant (Figures \ref{fig:Residence-probabilities-and}a
and \ref{fig:return_res_times_three_half}a). At low values of $\alpha$,
the system spends significant time exploring phase space away from
the eigenstates of $S_{z}$, but this behaviour becomes rarer as $\alpha$
increases: for example, the spin 3/2 residence probabilities for the
four eigenstates rise to about 0.25 for $\alpha>6$. Such a localisation
is also demonstrated by the pattern of stationary probability densities
over $\langle S_{z}\rangle$ and $\langle S_{y}\rangle$ for spin
1 (Figure \ref{fig:histograms_spin_1}) such that, for high values
of $\alpha$, the density is almost exclusively confined to regions
of phase space in the vicinity of the three eigenstates. The non-unitary
measurement dynamics appear to dominate, localising the system near
the eigenstates and supporting the conventional picture of wavefunction
collapse.

Notably, in the spin 1 and spin 3/2 systems, the QZE does not manifest
itself in the same way for each of the eigenstates of the corresponding
$S_{z}$ observable. Figures \ref{fig:Residence-probabilities-and}b
and \ref{fig:return_res_times_three_half}b demonstrate that there
is an asymmetry in the dwell and return behaviour for each of the
eigenstates of the measured observable. For the eigenstates at the
extremities (such as the $|\pm1\rangle$ eigenstates of the spin 1
system or the $|\pm\frac{3}{2}\rangle$ eigenstates of the spin 3/2
system), the mean return time is roughly double that of the middle
eigenstates (for example, the $|0\rangle$ eigenstate of the spin
1 system or the $|\pm\frac{1}{2}\rangle$ eigenstates of the spin
3/2 system) when $\alpha=7$, thus the QZE manifests itself more strongly
for the eigenstates at the extremities. In Bauer \textit{et al}, the
jump rates in a strong measurement regime can be calculated, with
the inverse yielding an average time between two jumps \citep{bauer2015a}.
Utilising their analytical result for the jump rates, we have found
that the average time between two jumps out of the spin 1 $|\pm1\rangle$
eigenstates or the $|\pm\frac{3}{2}\rangle$ eigenstates of the spin
3/2 system was roughly double that of the spin 1 $|0\rangle$ eigenstate
or the spin 3/2 $|\pm\frac{1}{2}\rangle$ eigenstates, confirming
our results for high measurement strength in Figures \ref{fig:Residence-probabilities-and}b
and \ref{fig:return_res_times_three_half}b. Moreover, the measurement
strength changes the eigenstates available to the spin 1 and spin
3/2 systems such that in Figures \ref{fig:spin1_trajectories} and
\ref{fig:spin_three_half_plots}, for a low measurement strength only
the $|\pm1\rangle$ or $|\pm\frac{3}{2}\rangle$ respectively are
frequently visited. Both of these features could be of use in quantum
state control protocols .

Aside from demonstrating the QZE in an open quantum spin system, the
QSD framework for generating stochastic quantum trajectories could
be used to shed light onto other features of quantum systems such
as entanglement, decoherence and measurement back-action. It could
also reveal behaviour in more complex systems such as those possessing
a memory of past environment-system interactions. Whilst the dynamics
of some quantum systems might be well approximated by Markovian stochastic
trajectories, future work could consider non-Markovian unravellings
or non-Markovian master equations generating stochastic trajectories
that are not constrained by the Born-Markov approximation, enabling
the study of complex environments with non-negligible correlation
times. 
\begin{acknowledgments}
SMW is supported by a PhD studentship funded by EPSRC under grant
codes EP/R513143/1 and EP/T517793/1.
\end{acknowledgments}

\bibliography{QZE_paper_better_bibtex2}

\begin{thebibliography}{67}%
\makeatletter
\providecommand \@ifxundefined [1]{%
 \@ifx{#1\undefined}
}%
\providecommand \@ifnum [1]{%
 \ifnum #1\expandafter \@firstoftwo
 \else \expandafter \@secondoftwo
 \fi
}%
\providecommand \@ifx [1]{%
 \ifx #1\expandafter \@firstoftwo
 \else \expandafter \@secondoftwo
 \fi
}%
\providecommand \natexlab [1]{#1}%
\providecommand \enquote  [1]{``#1''}%
\providecommand \bibnamefont  [1]{#1}%
\providecommand \bibfnamefont [1]{#1}%
\providecommand \citenamefont [1]{#1}%
\providecommand \href@noop [0]{\@secondoftwo}%
\providecommand \href [0]{\begingroup \@sanitize@url \@href}%
\providecommand \@href[1]{\@@startlink{#1}\@@href}%
\providecommand \@@href[1]{\endgroup#1\@@endlink}%
\providecommand \@sanitize@url [0]{\catcode `\\12\catcode `\$12\catcode `\&12\catcode `\#12\catcode `\^12\catcode `\_12\catcode `\%12\relax}%
\providecommand \@@startlink[1]{}%
\providecommand \@@endlink[0]{}%
\providecommand \url  [0]{\begingroup\@sanitize@url \@url }%
\providecommand \@url [1]{\endgroup\@href {#1}{\urlprefix }}%
\providecommand \urlprefix  [0]{URL }%
\providecommand \Eprint [0]{\href }%
\providecommand \doibase [0]{https://doi.org/}%
\providecommand \selectlanguage [0]{\@gobble}%
\providecommand \bibinfo  [0]{\@secondoftwo}%
\providecommand \bibfield  [0]{\@secondoftwo}%
\providecommand \translation [1]{[#1]}%
\providecommand \BibitemOpen [0]{}%
\providecommand \bibitemStop [0]{}%
\providecommand \bibitemNoStop [0]{.\EOS\space}%
\providecommand \EOS [0]{\spacefactor3000\relax}%
\providecommand \BibitemShut  [1]{\csname bibitem#1\endcsname}%
\let\auto@bib@innerbib\@empty
\bibitem [{\citenamefont {Degasperis}\ \emph {et~al.}(1974)\citenamefont {Degasperis}, \citenamefont {Fonda},\ and\ \citenamefont {Ghirardi}}]{degasperis1974}%
  \BibitemOpen
  \bibfield  {author} {\bibinfo {author} {\bibfnamefont {A.}~\bibnamefont {Degasperis}}, \bibinfo {author} {\bibfnamefont {L.}~\bibnamefont {Fonda}},\ and\ \bibinfo {author} {\bibfnamefont {G.~C.}\ \bibnamefont {Ghirardi}},\ }\bibfield  {title} {\bibinfo {title} {Does the lifetime of an unstable system depend on the measuring apparatus?},\ }\href@noop {} {\bibfield  {journal} {\bibinfo  {journal} {Il Nuovo Cimento A}\ }\textbf {\bibinfo {volume} {21}},\ \bibinfo {pages} {471} (\bibinfo {year} {1974})}\BibitemShut {NoStop}%
\bibitem [{\citenamefont {Misra}\ and\ \citenamefont {Sudarshan}(1977)}]{misra1977a}%
  \BibitemOpen
  \bibfield  {author} {\bibinfo {author} {\bibfnamefont {B.}~\bibnamefont {Misra}}\ and\ \bibinfo {author} {\bibfnamefont {E.~C.~G.}\ \bibnamefont {Sudarshan}},\ }\bibfield  {title} {\bibinfo {title} {The {{Zeno}}'s paradox in quantum theory},\ }\href {https://doi.org/10.1063/1.523304} {\bibfield  {journal} {\bibinfo  {journal} {Journal of Mathematical Physics}\ }\textbf {\bibinfo {volume} {18}},\ \bibinfo {pages} {756} (\bibinfo {year} {1977})}\BibitemShut {NoStop}%
\bibitem [{\citenamefont {Zhang}\ \emph {et~al.}(2019)\citenamefont {Zhang}, \citenamefont {Wu}, \citenamefont {Xie}, \citenamefont {Wu},\ and\ \citenamefont {Chen}}]{zhang2019}%
  \BibitemOpen
  \bibfield  {author} {\bibinfo {author} {\bibfnamefont {M.}~\bibnamefont {Zhang}}, \bibinfo {author} {\bibfnamefont {C.}~\bibnamefont {Wu}}, \bibinfo {author} {\bibfnamefont {Y.}~\bibnamefont {Xie}}, \bibinfo {author} {\bibfnamefont {W.}~\bibnamefont {Wu}},\ and\ \bibinfo {author} {\bibfnamefont {P.}~\bibnamefont {Chen}},\ }\bibfield  {title} {\bibinfo {title} {Quantum {{Zeno}} effect by incomplete measurements},\ }\href {https://doi.org/10.1007/s11128-019-2194-9} {\bibfield  {journal} {\bibinfo  {journal} {Quantum Information Processing}\ }\textbf {\bibinfo {volume} {18}},\ \bibinfo {pages} {97} (\bibinfo {year} {2019})}\BibitemShut {NoStop}%
\bibitem [{\citenamefont {Peres}\ and\ \citenamefont {Ron}(1990)}]{peres1990a}%
  \BibitemOpen
  \bibfield  {author} {\bibinfo {author} {\bibfnamefont {A.}~\bibnamefont {Peres}}\ and\ \bibinfo {author} {\bibfnamefont {A.}~\bibnamefont {Ron}},\ }\bibfield  {title} {\bibinfo {title} {Incomplete ``collapse'' and partial quantum {{Zeno}} effect},\ }\href {https://doi.org/10.1103/PhysRevA.42.5720} {\bibfield  {journal} {\bibinfo  {journal} {Physical Review A}\ }\textbf {\bibinfo {volume} {42}},\ \bibinfo {pages} {5720} (\bibinfo {year} {1990})}\BibitemShut {NoStop}%
\bibitem [{\citenamefont {Oliveira}\ \emph {et~al.}(2022)\citenamefont {Oliveira}, \citenamefont {Higgins}, \citenamefont {Zhang}, \citenamefont {Predojevi{\'c}}, \citenamefont {Hennrich}, \citenamefont {Bachelard},\ and\ \citenamefont {{Villas-Boas}}}]{oliveira2022}%
  \BibitemOpen
  \bibfield  {author} {\bibinfo {author} {\bibfnamefont {M.~H.}\ \bibnamefont {Oliveira}}, \bibinfo {author} {\bibfnamefont {G.}~\bibnamefont {Higgins}}, \bibinfo {author} {\bibfnamefont {C.}~\bibnamefont {Zhang}}, \bibinfo {author} {\bibfnamefont {A.}~\bibnamefont {Predojevi{\'c}}}, \bibinfo {author} {\bibfnamefont {M.}~\bibnamefont {Hennrich}}, \bibinfo {author} {\bibfnamefont {R.}~\bibnamefont {Bachelard}},\ and\ \bibinfo {author} {\bibfnamefont {C.~J.}\ \bibnamefont {{Villas-Boas}}},\ }\href {https://doi.org/10.48550/arXiv.2205.10590} {\bibinfo {title} {Steady-state entanglement generation for non-degenerate qubits}} (\bibinfo {year} {2022}),\ \Eprint {https://arxiv.org/abs/2205.10590} {arxiv:2205.10590 [quant-ph]} \BibitemShut {NoStop}%
\bibitem [{\citenamefont {Salih}\ \emph {et~al.}(2021)\citenamefont {Salih}, \citenamefont {Hance}, \citenamefont {McCutcheon}, \citenamefont {Rudolph},\ and\ \citenamefont {Rarity}}]{salih2021}%
  \BibitemOpen
  \bibfield  {author} {\bibinfo {author} {\bibfnamefont {H.}~\bibnamefont {Salih}}, \bibinfo {author} {\bibfnamefont {J.~R.}\ \bibnamefont {Hance}}, \bibinfo {author} {\bibfnamefont {W.}~\bibnamefont {McCutcheon}}, \bibinfo {author} {\bibfnamefont {T.}~\bibnamefont {Rudolph}},\ and\ \bibinfo {author} {\bibfnamefont {J.}~\bibnamefont {Rarity}},\ }\bibfield  {title} {\bibinfo {title} {Deterministic {{Teleportation}} and {{Universal Computation Without Particle Exchange}}},\ }\Eprint {https://arxiv.org/abs/2009.05564} {arxiv:2009.05564 [quant-ph]}  (\bibinfo {year} {2021})\BibitemShut {NoStop}%
\bibitem [{\citenamefont {Chen}\ and\ \citenamefont {Brun}(2020)}]{chen2020}%
  \BibitemOpen
  \bibfield  {author} {\bibinfo {author} {\bibfnamefont {Y.-H.}\ \bibnamefont {Chen}}\ and\ \bibinfo {author} {\bibfnamefont {T.~A.}\ \bibnamefont {Brun}},\ }\bibfield  {title} {\bibinfo {title} {Continuous quantum error detection and suppression with pairwise local interactions},\ }\href {https://doi.org/10.1103/PhysRevResearch.2.043093} {\bibfield  {journal} {\bibinfo  {journal} {Physical Review Research}\ }\textbf {\bibinfo {volume} {2}},\ \bibinfo {pages} {043093} (\bibinfo {year} {2020})}\BibitemShut {NoStop}%
\bibitem [{\citenamefont {Li}\ \emph {et~al.}(2018)\citenamefont {Li}, \citenamefont {Chen},\ and\ \citenamefont {Fisher}}]{li2018}%
  \BibitemOpen
  \bibfield  {author} {\bibinfo {author} {\bibfnamefont {Y.}~\bibnamefont {Li}}, \bibinfo {author} {\bibfnamefont {X.}~\bibnamefont {Chen}},\ and\ \bibinfo {author} {\bibfnamefont {M.~P.~A.}\ \bibnamefont {Fisher}},\ }\bibfield  {title} {\bibinfo {title} {Quantum {{Zeno Effect}} and the {{Many-body Entanglement Transition}}},\ }\href {https://doi.org/10.1103/PhysRevB.98.205136} {\bibfield  {journal} {\bibinfo  {journal} {Physical Review B}\ }\textbf {\bibinfo {volume} {98}},\ \bibinfo {pages} {205136} (\bibinfo {year} {2018})},\ \Eprint {https://arxiv.org/abs/1808.06134} {arxiv:1808.06134 [cond-mat, physics:quant-ph]} \BibitemShut {NoStop}%
\bibitem [{\citenamefont {Lin}\ \emph {et~al.}(2022)\citenamefont {Lin}, \citenamefont {Huang}, \citenamefont {Lambert}, \citenamefont {Chen}, \citenamefont {Nori},\ and\ \citenamefont {Chen}}]{lin2022}%
  \BibitemOpen
  \bibfield  {author} {\bibinfo {author} {\bibfnamefont {J.-D.}\ \bibnamefont {Lin}}, \bibinfo {author} {\bibfnamefont {C.-Y.}\ \bibnamefont {Huang}}, \bibinfo {author} {\bibfnamefont {N.}~\bibnamefont {Lambert}}, \bibinfo {author} {\bibfnamefont {G.-Y.}\ \bibnamefont {Chen}}, \bibinfo {author} {\bibfnamefont {F.}~\bibnamefont {Nori}},\ and\ \bibinfo {author} {\bibfnamefont {Y.-N.}\ \bibnamefont {Chen}},\ }\href {https://doi.org/10.48550/arXiv.2208.02472} {\bibinfo {title} {Space-time dual quantum {{Zeno}} effect: {{Interferometric}} engineering of open quantum system dynamics}} (\bibinfo {year} {2022}),\ \Eprint {https://arxiv.org/abs/2208.02472} {arxiv:2208.02472 [quant-ph]} \BibitemShut {NoStop}%
\bibitem [{\citenamefont {Long}\ \emph {et~al.}(2022)\citenamefont {Long}, \citenamefont {He}, \citenamefont {Zhang}, \citenamefont {Tang}, \citenamefont {Lin}, \citenamefont {Liu}, \citenamefont {Nie}, \citenamefont {Feng}, \citenamefont {Li}, \citenamefont {Xin}, \citenamefont {Ai},\ and\ \citenamefont {Lu}}]{long2022}%
  \BibitemOpen
  \bibfield  {author} {\bibinfo {author} {\bibfnamefont {X.}~\bibnamefont {Long}}, \bibinfo {author} {\bibfnamefont {W.-T.}\ \bibnamefont {He}}, \bibinfo {author} {\bibfnamefont {N.-N.}\ \bibnamefont {Zhang}}, \bibinfo {author} {\bibfnamefont {K.}~\bibnamefont {Tang}}, \bibinfo {author} {\bibfnamefont {Z.}~\bibnamefont {Lin}}, \bibinfo {author} {\bibfnamefont {H.}~\bibnamefont {Liu}}, \bibinfo {author} {\bibfnamefont {X.}~\bibnamefont {Nie}}, \bibinfo {author} {\bibfnamefont {G.}~\bibnamefont {Feng}}, \bibinfo {author} {\bibfnamefont {J.}~\bibnamefont {Li}}, \bibinfo {author} {\bibfnamefont {T.}~\bibnamefont {Xin}}, \bibinfo {author} {\bibfnamefont {Q.}~\bibnamefont {Ai}},\ and\ \bibinfo {author} {\bibfnamefont {D.}~\bibnamefont {Lu}},\ }\bibfield  {title} {\bibinfo {title} {Entanglement-{{Enhanced Quantum Metrology}} in {{Colored Noise}} by {{Quantum Zeno Effect}}},\ }\href {https://doi.org/10.1103/PhysRevLett.129.070502} {\bibfield  {journal} {\bibinfo  {journal} {Physical Review Letters}\ }\textbf
  {\bibinfo {volume} {129}},\ \bibinfo {pages} {070502} (\bibinfo {year} {2022})}\BibitemShut {NoStop}%
\bibitem [{\citenamefont {Blumenthal}\ \emph {et~al.}(2021)\citenamefont {Blumenthal}, \citenamefont {Mor}, \citenamefont {Diringer}, \citenamefont {Martin}, \citenamefont {Lewalle}, \citenamefont {Burgarth}, \citenamefont {Whaley},\ and\ \citenamefont {{Hacohen-Gourgy}}}]{blumenthal2021}%
  \BibitemOpen
  \bibfield  {author} {\bibinfo {author} {\bibfnamefont {E.}~\bibnamefont {Blumenthal}}, \bibinfo {author} {\bibfnamefont {C.}~\bibnamefont {Mor}}, \bibinfo {author} {\bibfnamefont {A.~A.}\ \bibnamefont {Diringer}}, \bibinfo {author} {\bibfnamefont {L.~S.}\ \bibnamefont {Martin}}, \bibinfo {author} {\bibfnamefont {P.}~\bibnamefont {Lewalle}}, \bibinfo {author} {\bibfnamefont {D.}~\bibnamefont {Burgarth}}, \bibinfo {author} {\bibfnamefont {K.~B.}\ \bibnamefont {Whaley}},\ and\ \bibinfo {author} {\bibfnamefont {S.}~\bibnamefont {{Hacohen-Gourgy}}},\ }\href {https://doi.org/10.48550/arXiv.2108.08549} {\bibinfo {title} {Demonstration of universal control between non-interacting qubits using the {{Quantum Zeno}} effect}} (\bibinfo {year} {2021}),\ \Eprint {https://arxiv.org/abs/2108.08549} {arxiv:2108.08549 [quant-ph]} \BibitemShut {NoStop}%
\bibitem [{\citenamefont {Nodurft}\ \emph {et~al.}(2022)\citenamefont {Nodurft}, \citenamefont {Shaw}, \citenamefont {Glasser}, \citenamefont {Kirby},\ and\ \citenamefont {Searles}}]{nodurft2022}%
  \BibitemOpen
  \bibfield  {author} {\bibinfo {author} {\bibfnamefont {I.~C.}\ \bibnamefont {Nodurft}}, \bibinfo {author} {\bibfnamefont {H.~C.}\ \bibnamefont {Shaw}}, \bibinfo {author} {\bibfnamefont {R.~T.}\ \bibnamefont {Glasser}}, \bibinfo {author} {\bibfnamefont {B.~T.}\ \bibnamefont {Kirby}},\ and\ \bibinfo {author} {\bibfnamefont {T.~A.}\ \bibnamefont {Searles}},\ }\href {https://doi.org/10.48550/arXiv.2205.02315} {\bibinfo {title} {Generation of polarization entanglement via the quantum {{Zeno}} effect}} (\bibinfo {year} {2022}),\ \Eprint {https://arxiv.org/abs/2205.02315} {arxiv:2205.02315 [quant-ph]} \BibitemShut {NoStop}%
\bibitem [{\citenamefont {Leppenen}\ and\ \citenamefont {Smirnov}(2022)}]{leppenen2022}%
  \BibitemOpen
  \bibfield  {author} {\bibinfo {author} {\bibfnamefont {N.~V.}\ \bibnamefont {Leppenen}}\ and\ \bibinfo {author} {\bibfnamefont {D.~S.}\ \bibnamefont {Smirnov}},\ }\href {https://doi.org/10.48550/arXiv.2202.13994} {\bibinfo {title} {Optical measurement of electron spins in quantum dots: {{Quantum Zeno}} effects}} (\bibinfo {year} {2022}),\ \Eprint {https://arxiv.org/abs/2202.13994} {arxiv:2202.13994 [cond-mat, physics:quant-ph]} \BibitemShut {NoStop}%
\bibitem [{\citenamefont {Patsch}\ \emph {et~al.}(2020)\citenamefont {Patsch}, \citenamefont {Maniscalco},\ and\ \citenamefont {Koch}}]{patsch2020}%
  \BibitemOpen
  \bibfield  {author} {\bibinfo {author} {\bibfnamefont {S.}~\bibnamefont {Patsch}}, \bibinfo {author} {\bibfnamefont {S.}~\bibnamefont {Maniscalco}},\ and\ \bibinfo {author} {\bibfnamefont {C.~P.}\ \bibnamefont {Koch}},\ }\bibfield  {title} {\bibinfo {title} {Simulation of open-quantum-system dynamics using the quantum {{Zeno}} effect},\ }\href {https://doi.org/10.1103/PhysRevResearch.2.023133} {\bibfield  {journal} {\bibinfo  {journal} {Physical Review Research}\ }\textbf {\bibinfo {volume} {2}},\ \bibinfo {pages} {023133} (\bibinfo {year} {2020})}\BibitemShut {NoStop}%
\bibitem [{\citenamefont {Bethke}\ \emph {et~al.}(2020)\citenamefont {Bethke}, \citenamefont {McNeil}, \citenamefont {Ritzmann}, \citenamefont {Botzem}, \citenamefont {Ludwig}, \citenamefont {Wieck},\ and\ \citenamefont {Bluhm}}]{bethke2020}%
  \BibitemOpen
  \bibfield  {author} {\bibinfo {author} {\bibfnamefont {P.}~\bibnamefont {Bethke}}, \bibinfo {author} {\bibfnamefont {R.~P.~G.}\ \bibnamefont {McNeil}}, \bibinfo {author} {\bibfnamefont {J.}~\bibnamefont {Ritzmann}}, \bibinfo {author} {\bibfnamefont {T.}~\bibnamefont {Botzem}}, \bibinfo {author} {\bibfnamefont {A.}~\bibnamefont {Ludwig}}, \bibinfo {author} {\bibfnamefont {A.~D.}\ \bibnamefont {Wieck}},\ and\ \bibinfo {author} {\bibfnamefont {H.}~\bibnamefont {Bluhm}},\ }\bibfield  {title} {\bibinfo {title} {Measurement of {{Backaction}} from {{Electron Spins}} in a {{Gate-Defined GaAs Double Quantum}} dot {{Coupled}} to a {{Mesoscopic Nuclear Spin Bath}}},\ }\href {https://doi.org/10.1103/PhysRevLett.125.047701} {\bibfield  {journal} {\bibinfo  {journal} {Physical Review Letters}\ }\textbf {\bibinfo {volume} {125}},\ \bibinfo {pages} {047701} (\bibinfo {year} {2020})}\BibitemShut {NoStop}%
\bibitem [{\citenamefont {M{\k{a}}dzik}\ \emph {et~al.}(2020)\citenamefont {M{\k{a}}dzik}, \citenamefont {Ladd}, \citenamefont {Hudson}, \citenamefont {Itoh}, \citenamefont {Jakob}, \citenamefont {Johnson}, \citenamefont {McCallum}, \citenamefont {Jamieson}, \citenamefont {Dzurak}, \citenamefont {Laucht},\ and\ \citenamefont {Morello}}]{madzik2020}%
  \BibitemOpen
  \bibfield  {author} {\bibinfo {author} {\bibfnamefont {M.~T.}\ \bibnamefont {M{\k{a}}dzik}}, \bibinfo {author} {\bibfnamefont {T.~D.}\ \bibnamefont {Ladd}}, \bibinfo {author} {\bibfnamefont {F.~E.}\ \bibnamefont {Hudson}}, \bibinfo {author} {\bibfnamefont {K.~M.}\ \bibnamefont {Itoh}}, \bibinfo {author} {\bibfnamefont {A.~M.}\ \bibnamefont {Jakob}}, \bibinfo {author} {\bibfnamefont {B.~C.}\ \bibnamefont {Johnson}}, \bibinfo {author} {\bibfnamefont {J.~C.}\ \bibnamefont {McCallum}}, \bibinfo {author} {\bibfnamefont {D.~N.}\ \bibnamefont {Jamieson}}, \bibinfo {author} {\bibfnamefont {A.~S.}\ \bibnamefont {Dzurak}}, \bibinfo {author} {\bibfnamefont {A.}~\bibnamefont {Laucht}},\ and\ \bibinfo {author} {\bibfnamefont {A.}~\bibnamefont {Morello}},\ }\bibfield  {title} {\bibinfo {title} {Controllable freezing of the nuclear spin bath in a single-atom spin qubit},\ }\href {https://doi.org/10.1126/sciadv.aba3442} {\bibfield  {journal} {\bibinfo  {journal} {Science Advances}\ }\textbf {\bibinfo {volume} {6}},\
  \bibinfo {pages} {eaba3442} (\bibinfo {year} {2020})}\BibitemShut {NoStop}%
\bibitem [{\citenamefont {Norsen}(2017)}]{norsen2017}%
  \BibitemOpen
  \bibfield  {author} {\bibinfo {author} {\bibfnamefont {T.}~\bibnamefont {Norsen}},\ }\href@noop {} {\emph {\bibinfo {title} {Foundations of {{Quantum Mechanics}}: {{An Exploration}} of the {{Physical Meaning}} of {{Quantum Theory}}}}}\ (\bibinfo  {publisher} {{Springer}},\ \bibinfo {address} {{Cham, Switzerland}},\ \bibinfo {year} {2017})\BibitemShut {NoStop}%
\bibitem [{\citenamefont {Jacobs}(2014)}]{jacobs2014a}%
  \BibitemOpen
  \bibfield  {author} {\bibinfo {author} {\bibfnamefont {K.}~\bibnamefont {Jacobs}},\ }\href {https://doi.org/10.1017/CBO9781139179027} {\emph {\bibinfo {title} {Quantum {{Measurement Theory}} and Its {{Applications}}}}}\ (\bibinfo  {publisher} {{Cambridge University Press}},\ \bibinfo {address} {{Cambridge}},\ \bibinfo {year} {2014})\BibitemShut {NoStop}%
\bibitem [{\citenamefont {Murch}\ \emph {et~al.}(2013)\citenamefont {Murch}, \citenamefont {Weber}, \citenamefont {Macklin},\ and\ \citenamefont {Siddiqi}}]{murch2013}%
  \BibitemOpen
  \bibfield  {author} {\bibinfo {author} {\bibfnamefont {K.~W.}\ \bibnamefont {Murch}}, \bibinfo {author} {\bibfnamefont {S.~J.}\ \bibnamefont {Weber}}, \bibinfo {author} {\bibfnamefont {C.}~\bibnamefont {Macklin}},\ and\ \bibinfo {author} {\bibfnamefont {I.}~\bibnamefont {Siddiqi}},\ }\bibfield  {title} {\bibinfo {title} {Observing single quantum trajectories of a superconducting quantum bit},\ }\href {https://doi.org/10.1038/nature12539} {\bibfield  {journal} {\bibinfo  {journal} {Nature}\ }\textbf {\bibinfo {volume} {502}},\ \bibinfo {pages} {211} (\bibinfo {year} {2013})}\BibitemShut {NoStop}%
\bibitem [{\citenamefont {Jordan}(2013)}]{jordan2013}%
  \BibitemOpen
  \bibfield  {author} {\bibinfo {author} {\bibfnamefont {A.~N.}\ \bibnamefont {Jordan}},\ }\bibfield  {title} {\bibinfo {title} {Watching the wavefunction collapse},\ }\href {https://doi.org/10.1038/502177a} {\bibfield  {journal} {\bibinfo  {journal} {Nature}\ }\textbf {\bibinfo {volume} {502}},\ \bibinfo {pages} {177} (\bibinfo {year} {2013})}\BibitemShut {NoStop}%
\bibitem [{\citenamefont {Bellini}\ \emph {et~al.}(2022)\citenamefont {Bellini}, \citenamefont {Kwon}, \citenamefont {Biagi}, \citenamefont {Francesconi}, \citenamefont {Zavatta},\ and\ \citenamefont {Kim}}]{bellini2022}%
  \BibitemOpen
  \bibfield  {author} {\bibinfo {author} {\bibfnamefont {M.}~\bibnamefont {Bellini}}, \bibinfo {author} {\bibfnamefont {H.}~\bibnamefont {Kwon}}, \bibinfo {author} {\bibfnamefont {N.}~\bibnamefont {Biagi}}, \bibinfo {author} {\bibfnamefont {S.}~\bibnamefont {Francesconi}}, \bibinfo {author} {\bibfnamefont {A.}~\bibnamefont {Zavatta}},\ and\ \bibinfo {author} {\bibfnamefont {M.~S.}\ \bibnamefont {Kim}},\ }\href {https://doi.org/10.48550/arXiv.2205.13089} {\bibinfo {title} {Demonstrating quantum microscopic reversibility using coherent states of light}} (\bibinfo {year} {2022}),\ \Eprint {https://arxiv.org/abs/2205.13089} {arxiv:2205.13089 [quant-ph]} \BibitemShut {NoStop}%
\bibitem [{\citenamefont {Gross}\ \emph {et~al.}(2022)\citenamefont {Gross}, \citenamefont {Baragiola}, \citenamefont {Stace},\ and\ \citenamefont {Combes}}]{gross2022}%
  \BibitemOpen
  \bibfield  {author} {\bibinfo {author} {\bibfnamefont {J.~A.}\ \bibnamefont {Gross}}, \bibinfo {author} {\bibfnamefont {B.~Q.}\ \bibnamefont {Baragiola}}, \bibinfo {author} {\bibfnamefont {T.~M.}\ \bibnamefont {Stace}},\ and\ \bibinfo {author} {\bibfnamefont {J.}~\bibnamefont {Combes}},\ }\bibfield  {title} {\bibinfo {title} {Master equations and quantum trajectories for squeezed wave packets},\ }\href {https://doi.org/10.1103/PhysRevA.105.023721} {\bibfield  {journal} {\bibinfo  {journal} {Physical Review A}\ }\textbf {\bibinfo {volume} {105}},\ \bibinfo {pages} {023721} (\bibinfo {year} {2022})}\BibitemShut {NoStop}%
\bibitem [{\citenamefont {Minev}\ \emph {et~al.}(2019)\citenamefont {Minev}, \citenamefont {Mundhada}, \citenamefont {Shankar}, \citenamefont {Reinhold}, \citenamefont {{Guti{\'e}rrez-J{\'a}uregui}}, \citenamefont {Schoelkopf}, \citenamefont {Mirrahimi}, \citenamefont {Carmichael},\ and\ \citenamefont {Devoret}}]{minev2019}%
  \BibitemOpen
  \bibfield  {author} {\bibinfo {author} {\bibfnamefont {Z.~K.}\ \bibnamefont {Minev}}, \bibinfo {author} {\bibfnamefont {S.~O.}\ \bibnamefont {Mundhada}}, \bibinfo {author} {\bibfnamefont {S.}~\bibnamefont {Shankar}}, \bibinfo {author} {\bibfnamefont {P.}~\bibnamefont {Reinhold}}, \bibinfo {author} {\bibfnamefont {R.}~\bibnamefont {{Guti{\'e}rrez-J{\'a}uregui}}}, \bibinfo {author} {\bibfnamefont {R.~J.}\ \bibnamefont {Schoelkopf}}, \bibinfo {author} {\bibfnamefont {M.}~\bibnamefont {Mirrahimi}}, \bibinfo {author} {\bibfnamefont {H.~J.}\ \bibnamefont {Carmichael}},\ and\ \bibinfo {author} {\bibfnamefont {M.~H.}\ \bibnamefont {Devoret}},\ }\bibfield  {title} {\bibinfo {title} {To catch and reverse a quantum jump mid-flight},\ }\href {https://doi.org/10.1038/s41586-019-1287-z} {\bibfield  {journal} {\bibinfo  {journal} {Nature}\ }\textbf {\bibinfo {volume} {570}},\ \bibinfo {pages} {200} (\bibinfo {year} {2019})}\BibitemShut {NoStop}%
\bibitem [{\citenamefont {Presilla}\ \emph {et~al.}(1996)\citenamefont {Presilla}, \citenamefont {Onofrio},\ and\ \citenamefont {Tambini}}]{presilla1996}%
  \BibitemOpen
  \bibfield  {author} {\bibinfo {author} {\bibfnamefont {C.}~\bibnamefont {Presilla}}, \bibinfo {author} {\bibfnamefont {R.}~\bibnamefont {Onofrio}},\ and\ \bibinfo {author} {\bibfnamefont {U.}~\bibnamefont {Tambini}},\ }\bibfield  {title} {\bibinfo {title} {Measurement {{Quantum Mechanics}} and {{Experiments}} on {{Quantum Zeno Effect}}},\ }\href {https://doi.org/10.1006/aphy.1996.0052} {\bibfield  {journal} {\bibinfo  {journal} {Annals of Physics}\ }\textbf {\bibinfo {volume} {248}},\ \bibinfo {pages} {95} (\bibinfo {year} {1996})}\BibitemShut {NoStop}%
\bibitem [{\citenamefont {Itano}\ \emph {et~al.}(1990)\citenamefont {Itano}, \citenamefont {Heinzen}, \citenamefont {Bollinger},\ and\ \citenamefont {Wineland}}]{itano1990}%
  \BibitemOpen
  \bibfield  {author} {\bibinfo {author} {\bibfnamefont {W.~M.}\ \bibnamefont {Itano}}, \bibinfo {author} {\bibfnamefont {D.~J.}\ \bibnamefont {Heinzen}}, \bibinfo {author} {\bibfnamefont {J.~J.}\ \bibnamefont {Bollinger}},\ and\ \bibinfo {author} {\bibfnamefont {D.~J.}\ \bibnamefont {Wineland}},\ }\bibfield  {title} {\bibinfo {title} {Quantum {{Zeno}} effect},\ }\href {https://doi.org/10.1103/PhysRevA.41.2295} {\bibfield  {journal} {\bibinfo  {journal} {Physical Review A}\ }\textbf {\bibinfo {volume} {41}},\ \bibinfo {pages} {2295} (\bibinfo {year} {1990})}\BibitemShut {NoStop}%
\bibitem [{\citenamefont {Snizhko}\ \emph {et~al.}(2020)\citenamefont {Snizhko}, \citenamefont {Kumar},\ and\ \citenamefont {Romito}}]{snizhko2020}%
  \BibitemOpen
  \bibfield  {author} {\bibinfo {author} {\bibfnamefont {K.}~\bibnamefont {Snizhko}}, \bibinfo {author} {\bibfnamefont {P.}~\bibnamefont {Kumar}},\ and\ \bibinfo {author} {\bibfnamefont {A.}~\bibnamefont {Romito}},\ }\bibfield  {title} {\bibinfo {title} {Quantum {{Zeno}} effect appears in stages},\ }\href {https://doi.org/10.1103/PhysRevResearch.2.033512} {\bibfield  {journal} {\bibinfo  {journal} {Physical Review Research}\ }\textbf {\bibinfo {volume} {2}},\ \bibinfo {pages} {033512} (\bibinfo {year} {2020})}\BibitemShut {NoStop}%
\bibitem [{\citenamefont {Gambetta}\ \emph {et~al.}(2008)\citenamefont {Gambetta}, \citenamefont {Blais}, \citenamefont {Boissonneault}, \citenamefont {Houck}, \citenamefont {Schuster},\ and\ \citenamefont {Girvin}}]{gambetta2008}%
  \BibitemOpen
  \bibfield  {author} {\bibinfo {author} {\bibfnamefont {J.}~\bibnamefont {Gambetta}}, \bibinfo {author} {\bibfnamefont {A.}~\bibnamefont {Blais}}, \bibinfo {author} {\bibfnamefont {M.}~\bibnamefont {Boissonneault}}, \bibinfo {author} {\bibfnamefont {A.~A.}\ \bibnamefont {Houck}}, \bibinfo {author} {\bibfnamefont {D.~I.}\ \bibnamefont {Schuster}},\ and\ \bibinfo {author} {\bibfnamefont {S.~M.}\ \bibnamefont {Girvin}},\ }\bibfield  {title} {\bibinfo {title} {Quantum trajectory approach to circuit {{QED}}: {{Quantum}} jumps and the {{Zeno}} effect},\ }\href {https://doi.org/10.1103/PhysRevA.77.012112} {\bibfield  {journal} {\bibinfo  {journal} {Physical Review A}\ }\textbf {\bibinfo {volume} {77}},\ \bibinfo {pages} {012112} (\bibinfo {year} {2008})}\BibitemShut {NoStop}%
\bibitem [{\citenamefont {{de Vega}}\ and\ \citenamefont {Alonso}(2017)}]{devega2017}%
  \BibitemOpen
  \bibfield  {author} {\bibinfo {author} {\bibfnamefont {I.}~\bibnamefont {{de Vega}}}\ and\ \bibinfo {author} {\bibfnamefont {D.}~\bibnamefont {Alonso}},\ }\bibfield  {title} {\bibinfo {title} {Dynamics of non-{{Markovian}} open quantum systems},\ }\href {https://doi.org/10.1103/RevModPhys.89.015001} {\bibfield  {journal} {\bibinfo  {journal} {Reviews of Modern Physics}\ }\textbf {\bibinfo {volume} {89}},\ \bibinfo {pages} {015001} (\bibinfo {year} {2017})}\BibitemShut {NoStop}%
\bibitem [{\citenamefont {Christie}\ \emph {et~al.}(2022)\citenamefont {Christie}, \citenamefont {Eastman}, \citenamefont {Schubert},\ and\ \citenamefont {Graefe}}]{christie2022}%
  \BibitemOpen
  \bibfield  {author} {\bibinfo {author} {\bibfnamefont {R.}~\bibnamefont {Christie}}, \bibinfo {author} {\bibfnamefont {J.}~\bibnamefont {Eastman}}, \bibinfo {author} {\bibfnamefont {R.}~\bibnamefont {Schubert}},\ and\ \bibinfo {author} {\bibfnamefont {E.-M.}\ \bibnamefont {Graefe}},\ }\bibfield  {title} {\bibinfo {title} {Quantum-jump vs stochastic {{Schr\"odinger}} dynamics for {{Gaussian}} states with quadratic {{Hamiltonians}} and linear {{Lindbladians}}},\ }\href {https://doi.org/10.1088/1751-8121/ac9d73} {\bibfield  {journal} {\bibinfo  {journal} {Journal of Physics A: Mathematical and Theoretical}\ }\textbf {\bibinfo {volume} {55}},\ \bibinfo {pages} {455302} (\bibinfo {year} {2022})}\BibitemShut {NoStop}%
\bibitem [{\citenamefont {Gardiner}\ \emph {et~al.}(1992)\citenamefont {Gardiner}, \citenamefont {Parkins},\ and\ \citenamefont {Zoller}}]{gardiner1992}%
  \BibitemOpen
  \bibfield  {author} {\bibinfo {author} {\bibfnamefont {C.~W.}\ \bibnamefont {Gardiner}}, \bibinfo {author} {\bibfnamefont {A.~S.}\ \bibnamefont {Parkins}},\ and\ \bibinfo {author} {\bibfnamefont {P.}~\bibnamefont {Zoller}},\ }\bibfield  {title} {\bibinfo {title} {Wave-function quantum stochastic differential equations and quantum-jump simulation methods},\ }\href {https://doi.org/10.1103/PhysRevA.46.4363} {\bibfield  {journal} {\bibinfo  {journal} {Physical Review A}\ }\textbf {\bibinfo {volume} {46}},\ \bibinfo {pages} {4363} (\bibinfo {year} {1992})}\BibitemShut {NoStop}%
\bibitem [{\citenamefont {Gneiting}\ \emph {et~al.}(2021)\citenamefont {Gneiting}, \citenamefont {Rozhkov},\ and\ \citenamefont {Nori}}]{gneiting2021}%
  \BibitemOpen
  \bibfield  {author} {\bibinfo {author} {\bibfnamefont {C.}~\bibnamefont {Gneiting}}, \bibinfo {author} {\bibfnamefont {A.~V.}\ \bibnamefont {Rozhkov}},\ and\ \bibinfo {author} {\bibfnamefont {F.}~\bibnamefont {Nori}},\ }\bibfield  {title} {\bibinfo {title} {Jump-time unraveling of {{Markovian}} open quantum systems},\ }\href {https://doi.org/10.1103/PhysRevA.104.062212} {\bibfield  {journal} {\bibinfo  {journal} {Physical Review A}\ }\textbf {\bibinfo {volume} {104}},\ \bibinfo {pages} {062212} (\bibinfo {year} {2021})}\BibitemShut {NoStop}%
\bibitem [{\citenamefont {Percival}(1998)}]{percival1998}%
  \BibitemOpen
  \bibfield  {author} {\bibinfo {author} {\bibfnamefont {I.}~\bibnamefont {Percival}},\ }\href@noop {} {\emph {\bibinfo {title} {Quantum {{State Diffusion}}}}}\ (\bibinfo  {publisher} {{Cambridge University Press}},\ \bibinfo {address} {{Cambridge}},\ \bibinfo {year} {1998})\BibitemShut {NoStop}%
\bibitem [{\citenamefont {Bauer}\ \emph {et~al.}(2015)\citenamefont {Bauer}, \citenamefont {Bernard},\ and\ \citenamefont {Tilloy}}]{bauer2015a}%
  \BibitemOpen
  \bibfield  {author} {\bibinfo {author} {\bibfnamefont {M.}~\bibnamefont {Bauer}}, \bibinfo {author} {\bibfnamefont {D.}~\bibnamefont {Bernard}},\ and\ \bibinfo {author} {\bibfnamefont {A.}~\bibnamefont {Tilloy}},\ }\bibfield  {title} {\bibinfo {title} {Computing the rates of measurement-induced quantum jumps},\ }\href {https://doi.org/10.1088/1751-8113/48/25/25FT02} {\bibfield  {journal} {\bibinfo  {journal} {Journal of Physics A: Mathematical and Theoretical}\ }\textbf {\bibinfo {volume} {48}},\ \bibinfo {pages} {25FT02} (\bibinfo {year} {2015})}\BibitemShut {NoStop}%
\bibitem [{\citenamefont {Spiller}(1994)}]{spiller1994}%
  \BibitemOpen
  \bibfield  {author} {\bibinfo {author} {\bibfnamefont {T.~P.}\ \bibnamefont {Spiller}},\ }\bibfield  {title} {\bibinfo {title} {The {{Zeno}} effect: Measurement versus decoherence},\ }\href {https://doi.org/10.1016/0375-9601(94)90238-0} {\bibfield  {journal} {\bibinfo  {journal} {Physics Letters A}\ }\textbf {\bibinfo {volume} {192}},\ \bibinfo {pages} {163} (\bibinfo {year} {1994})}\BibitemShut {NoStop}%
\bibitem [{\citenamefont {Penrose}(2014)}]{penrose2014a}%
  \BibitemOpen
  \bibfield  {author} {\bibinfo {author} {\bibfnamefont {R.}~\bibnamefont {Penrose}},\ }\bibfield  {title} {\bibinfo {title} {On the {{Gravitization}} of {{Quantum Mechanics}} 1: {{Quantum State Reduction}}},\ }\href {https://doi.org/10.1007/s10701-013-9770-0} {\bibfield  {journal} {\bibinfo  {journal} {Foundations of Physics}\ }\textbf {\bibinfo {volume} {44}},\ \bibinfo {pages} {557} (\bibinfo {year} {2014})}\BibitemShut {NoStop}%
\bibitem [{\citenamefont {Ghirardi}\ \emph {et~al.}(1986)\citenamefont {Ghirardi}, \citenamefont {Rimini},\ and\ \citenamefont {Weber}}]{ghirardi1986a}%
  \BibitemOpen
  \bibfield  {author} {\bibinfo {author} {\bibfnamefont {G.~C.}\ \bibnamefont {Ghirardi}}, \bibinfo {author} {\bibfnamefont {A.}~\bibnamefont {Rimini}},\ and\ \bibinfo {author} {\bibfnamefont {T.}~\bibnamefont {Weber}},\ }\bibfield  {title} {\bibinfo {title} {Unified dynamics for microscopic and macroscopic systems},\ }\href {https://doi.org/10.1103/PhysRevD.34.470} {\bibfield  {journal} {\bibinfo  {journal} {Physical Review D}\ }\textbf {\bibinfo {volume} {34}},\ \bibinfo {pages} {470} (\bibinfo {year} {1986})}\BibitemShut {NoStop}%
\bibitem [{\citenamefont {Bohm}(1952)}]{bohm1952b}%
  \BibitemOpen
  \bibfield  {author} {\bibinfo {author} {\bibfnamefont {D.}~\bibnamefont {Bohm}},\ }\bibfield  {title} {\bibinfo {title} {A {{Suggested Interpretation}} of the {{Quantum Theory}} in {{Terms}} of "{{Hidden}}" {{Variables}}. {{I}}},\ }\href {https://doi.org/10.1103/PhysRev.85.166} {\bibfield  {journal} {\bibinfo  {journal} {Physical Review}\ }\textbf {\bibinfo {volume} {85}},\ \bibinfo {pages} {166} (\bibinfo {year} {1952})}\BibitemShut {NoStop}%
\bibitem [{\citenamefont {Saunders}\ \emph {et~al.}(2010)\citenamefont {Saunders}, \citenamefont {Barrett}, \citenamefont {Kent},\ and\ \citenamefont {Wallace}}]{saunders2010}%
  \BibitemOpen
  \bibfield  {author} {\bibinfo {author} {\bibfnamefont {S.}~\bibnamefont {Saunders}}, \bibinfo {author} {\bibfnamefont {J.}~\bibnamefont {Barrett}}, \bibinfo {author} {\bibfnamefont {A.}~\bibnamefont {Kent}},\ and\ \bibinfo {author} {\bibfnamefont {D.}~\bibnamefont {Wallace}},\ }\href@noop {} {\emph {\bibinfo {title} {Many {{Worlds}}?: {{Everett}}, {{Quantum Theory}}, \& {{Reality}}}}}\ (\bibinfo  {publisher} {{OUP Oxford}},\ \bibinfo {year} {2010})\BibitemShut {NoStop}%
\bibitem [{\citenamefont {Everett}(1957)}]{everett1957}%
  \BibitemOpen
  \bibfield  {author} {\bibinfo {author} {\bibfnamefont {H.}~\bibnamefont {Everett}},\ }\bibfield  {title} {\bibinfo {title} {"{{Relative State}}" {{Formulation}} of {{Quantum Mechanics}}},\ }\href {https://doi.org/10.1103/RevModPhys.29.454} {\bibfield  {journal} {\bibinfo  {journal} {Reviews of Modern Physics}\ }\textbf {\bibinfo {volume} {29}},\ \bibinfo {pages} {454} (\bibinfo {year} {1957})}\BibitemShut {NoStop}%
\bibitem [{\citenamefont {Bassi}\ \emph {et~al.}(2013)\citenamefont {Bassi}, \citenamefont {Lochan}, \citenamefont {Satin}, \citenamefont {Singh},\ and\ \citenamefont {Ulbricht}}]{bassi2013}%
  \BibitemOpen
  \bibfield  {author} {\bibinfo {author} {\bibfnamefont {A.}~\bibnamefont {Bassi}}, \bibinfo {author} {\bibfnamefont {K.}~\bibnamefont {Lochan}}, \bibinfo {author} {\bibfnamefont {S.}~\bibnamefont {Satin}}, \bibinfo {author} {\bibfnamefont {T.~P.}\ \bibnamefont {Singh}},\ and\ \bibinfo {author} {\bibfnamefont {H.}~\bibnamefont {Ulbricht}},\ }\bibfield  {title} {\bibinfo {title} {Models of wave-function collapse, underlying theories, and experimental tests},\ }\href {https://doi.org/10.1103/RevModPhys.85.471} {\bibfield  {journal} {\bibinfo  {journal} {Reviews of Modern Physics}\ }\textbf {\bibinfo {volume} {85}},\ \bibinfo {pages} {471} (\bibinfo {year} {2013})}\BibitemShut {NoStop}%
\bibitem [{\citenamefont {Pusey}\ \emph {et~al.}(2012)\citenamefont {Pusey}, \citenamefont {Barrett},\ and\ \citenamefont {Rudolph}}]{pusey2012}%
  \BibitemOpen
  \bibfield  {author} {\bibinfo {author} {\bibfnamefont {M.~F.}\ \bibnamefont {Pusey}}, \bibinfo {author} {\bibfnamefont {J.}~\bibnamefont {Barrett}},\ and\ \bibinfo {author} {\bibfnamefont {T.}~\bibnamefont {Rudolph}},\ }\bibfield  {title} {\bibinfo {title} {On the reality of the quantum state},\ }\href {https://doi.org/10.1038/nphys2309} {\bibfield  {journal} {\bibinfo  {journal} {Nature Physics}\ }\textbf {\bibinfo {volume} {8}},\ \bibinfo {pages} {475} (\bibinfo {year} {2012})}\BibitemShut {NoStop}%
\bibitem [{\citenamefont {Lombardi}\ and\ \citenamefont {Dieks}(2021)}]{lombardi2021a}%
  \BibitemOpen
  \bibfield  {author} {\bibinfo {author} {\bibfnamefont {O.}~\bibnamefont {Lombardi}}\ and\ \bibinfo {author} {\bibfnamefont {D.}~\bibnamefont {Dieks}},\ }\bibfield  {title} {\bibinfo {title} {Modal {{Interpretations}} of {{Quantum Mechanics}}},\ }in\ \href@noop {} {\emph {\bibinfo {booktitle} {The {{Stanford Encyclopedia}} of {{Philosophy}}}}},\ \bibinfo {editor} {edited by\ \bibinfo {editor} {\bibfnamefont {E.~N.}\ \bibnamefont {Zalta}}}\ (\bibinfo  {publisher} {{Metaphysics Research Lab, Stanford University}},\ \bibinfo {year} {2021})\ \bibinfo {edition} {winter 2021}\ ed.\BibitemShut {Stop}%
\bibitem [{\citenamefont {Petersen}(1963)}]{petersen1963}%
  \BibitemOpen
  \bibfield  {author} {\bibinfo {author} {\bibfnamefont {A.}~\bibnamefont {Petersen}},\ }\bibfield  {title} {\bibinfo {title} {The {{Philosophy}} of {{Niels Bohr}}},\ }\href {https://doi.org/10.1080/00963402.1963.11454520} {\bibfield  {journal} {\bibinfo  {journal} {Bulletin of the Atomic Scientists}\ }\textbf {\bibinfo {volume} {19}},\ \bibinfo {pages} {8} (\bibinfo {year} {1963})}\BibitemShut {NoStop}%
\bibitem [{\citenamefont {Fuchs}\ \emph {et~al.}(2014)\citenamefont {Fuchs}, \citenamefont {Mermin},\ and\ \citenamefont {Schack}}]{fuchs2014}%
  \BibitemOpen
  \bibfield  {author} {\bibinfo {author} {\bibfnamefont {C.~A.}\ \bibnamefont {Fuchs}}, \bibinfo {author} {\bibfnamefont {N.~D.}\ \bibnamefont {Mermin}},\ and\ \bibinfo {author} {\bibfnamefont {R.}~\bibnamefont {Schack}},\ }\bibfield  {title} {\bibinfo {title} {An introduction to {{QBism}} with an application to the locality of quantum mechanics},\ }\href {https://doi.org/10.1119/1.4874855} {\bibfield  {journal} {\bibinfo  {journal} {American Journal of Physics}\ }\textbf {\bibinfo {volume} {82}},\ \bibinfo {pages} {749} (\bibinfo {year} {2014})}\BibitemShut {NoStop}%
\bibitem [{\citenamefont {Pitowsky}(2005)}]{pitowsky2005}%
  \BibitemOpen
  \bibfield  {author} {\bibinfo {author} {\bibfnamefont {I.}~\bibnamefont {Pitowsky}},\ }\href {https://doi.org/10.48550/arXiv.quant-ph/0510095} {\bibinfo {title} {Quantum mechanics as a theory of probability}} (\bibinfo {year} {2005}),\ \Eprint {https://arxiv.org/abs/quant-ph/0510095} {arxiv:quant-ph/0510095} \BibitemShut {NoStop}%
\bibitem [{\citenamefont {Healey}(2012)}]{healey2012}%
  \BibitemOpen
  \bibfield  {author} {\bibinfo {author} {\bibfnamefont {R.}~\bibnamefont {Healey}},\ }\bibfield  {title} {\bibinfo {title} {Quantum {{Theory}}: {{A Pragmatist Approach}}},\ }\href {https://doi.org/10.1093/bjps/axr054} {\bibfield  {journal} {\bibinfo  {journal} {The British Journal for the Philosophy of Science}\ }\textbf {\bibinfo {volume} {63}},\ \bibinfo {pages} {729} (\bibinfo {year} {2012})}\BibitemShut {NoStop}%
\bibitem [{\citenamefont {Harrigan}\ and\ \citenamefont {Spekkens}(2010)}]{harrigan2010a}%
  \BibitemOpen
  \bibfield  {author} {\bibinfo {author} {\bibfnamefont {N.}~\bibnamefont {Harrigan}}\ and\ \bibinfo {author} {\bibfnamefont {R.~W.}\ \bibnamefont {Spekkens}},\ }\bibfield  {title} {\bibinfo {title} {Einstein, {{Incompleteness}}, and the {{Epistemic View}} of~{{Quantum States}}},\ }\href {https://doi.org/10.1007/s10701-009-9347-0} {\bibfield  {journal} {\bibinfo  {journal} {Foundations of Physics}\ }\textbf {\bibinfo {volume} {40}},\ \bibinfo {pages} {125} (\bibinfo {year} {2010})}\BibitemShut {NoStop}%
\bibitem [{\citenamefont {Kochen}\ and\ \citenamefont {Specker}(1967)}]{kochen1967}%
  \BibitemOpen
  \bibfield  {author} {\bibinfo {author} {\bibfnamefont {S.}~\bibnamefont {Kochen}}\ and\ \bibinfo {author} {\bibfnamefont {E.~P.}\ \bibnamefont {Specker}},\ }\bibfield  {title} {\bibinfo {title} {The {{Problem}} of {{Hidden Variables}} in {{Quantum Mechanics}}},\ }\href@noop {} {\bibfield  {journal} {\bibinfo  {journal} {Journal of Mathematics and Mechanics}\ }\textbf {\bibinfo {volume} {17}},\ \bibinfo {pages} {59} (\bibinfo {year} {1967})},\ \Eprint {https://arxiv.org/abs/24902153} {24902153} \BibitemShut {NoStop}%
\bibitem [{\citenamefont {Brukner}\ and\ \citenamefont {Zeilinger}(2002)}]{brukner2002}%
  \BibitemOpen
  \bibfield  {author} {\bibinfo {author} {\bibfnamefont {C.}~\bibnamefont {Brukner}}\ and\ \bibinfo {author} {\bibfnamefont {A.}~\bibnamefont {Zeilinger}},\ }\href {https://doi.org/10.48550/arXiv.quant-ph/0212084} {\bibinfo {title} {Information and fundamental elements of the structure of quantum theory}} (\bibinfo {year} {2002}),\ \Eprint {https://arxiv.org/abs/quant-ph/0212084} {arxiv:quant-ph/0212084} \BibitemShut {NoStop}%
\bibitem [{\citenamefont {Mermin}(2013)}]{mermin2013}%
  \BibitemOpen
  \bibfield  {author} {\bibinfo {author} {\bibfnamefont {N.~D.}\ \bibnamefont {Mermin}},\ }\href {https://doi.org/10.48550/arXiv.1301.6551} {\bibinfo {title} {Annotated {{Interview}} with a {{QBist}} in the {{Making}}}} (\bibinfo {year} {2013}),\ \Eprint {https://arxiv.org/abs/1301.6551} {arxiv:1301.6551 [physics, physics:quant-ph]} \BibitemShut {NoStop}%
\bibitem [{\citenamefont {Rovelli}(2021)}]{rovelli2021}%
  \BibitemOpen
  \bibfield  {author} {\bibinfo {author} {\bibfnamefont {C.}~\bibnamefont {Rovelli}},\ }\href@noop {} {\emph {\bibinfo {title} {Helgoland}}}\ (\bibinfo  {publisher} {{Penguin Books}},\ \bibinfo {address} {{London}},\ \bibinfo {year} {2021})\BibitemShut {NoStop}%
\bibitem [{\citenamefont {Maudlin}(2019)}]{maudlin2019}%
  \BibitemOpen
  \bibfield  {author} {\bibinfo {author} {\bibfnamefont {T.}~\bibnamefont {Maudlin}},\ }\href@noop {} {\emph {\bibinfo {title} {Philosophy of Physics: Quantum Theory}}},\ Princeton Foundations of Contemporary Philosophy\ (\bibinfo  {publisher} {{Princeton University Press}},\ \bibinfo {address} {{Princeton}},\ \bibinfo {year} {2019})\BibitemShut {NoStop}%
\bibitem [{\citenamefont {Leifer}(2014)}]{leifer2014}%
  \BibitemOpen
  \bibfield  {author} {\bibinfo {author} {\bibfnamefont {M.~S.}\ \bibnamefont {Leifer}},\ }\bibfield  {title} {\bibinfo {title} {Is the quantum state real? {{An}} extended review of {$\psi$}-ontology theorems},\ }\href {https://doi.org/10.12743/quanta.v3i1.22} {\bibfield  {journal} {\bibinfo  {journal} {Quanta}\ }\textbf {\bibinfo {volume} {3}},\ \bibinfo {pages} {67} (\bibinfo {year} {2014})},\ \Eprint {https://arxiv.org/abs/1409.1570} {arxiv:1409.1570} \BibitemShut {NoStop}%
\bibitem [{\citenamefont {Hiley}\ \emph {et~al.}(2000)\citenamefont {Hiley}, \citenamefont {Callaghan},\ and\ \citenamefont {Maroney}}]{hiley2000}%
  \BibitemOpen
  \bibfield  {author} {\bibinfo {author} {\bibfnamefont {B.~J.}\ \bibnamefont {Hiley}}, \bibinfo {author} {\bibfnamefont {R.~E.}\ \bibnamefont {Callaghan}},\ and\ \bibinfo {author} {\bibfnamefont {O.}~\bibnamefont {Maroney}},\ }\href {https://doi.org/10.48550/arXiv.quant-ph/0010020} {\bibinfo {title} {Quantum trajectories, real, surreal or an approximation to a deeper process?}} (\bibinfo {year} {2000}),\ \Eprint {https://arxiv.org/abs/quant-ph/0010020} {arxiv:quant-ph/0010020} \BibitemShut {NoStop}%
\bibitem [{\citenamefont {Wiseman}(1996)}]{wiseman1996}%
  \BibitemOpen
  \bibfield  {author} {\bibinfo {author} {\bibfnamefont {H.~M.}\ \bibnamefont {Wiseman}},\ }\bibfield  {title} {\bibinfo {title} {Quantum trajectories and quantum measurement theory},\ }\href {https://doi.org/10.1088/1355-5111/8/1/015} {\bibfield  {journal} {\bibinfo  {journal} {Quantum and Semiclassical Optics: Journal of the European Optical Society Part B}\ }\textbf {\bibinfo {volume} {8}},\ \bibinfo {pages} {205} (\bibinfo {year} {1996})}\BibitemShut {NoStop}%
\bibitem [{\citenamefont {Gambetta}\ and\ \citenamefont {Wiseman}(2003)}]{gambetta2003}%
  \BibitemOpen
  \bibfield  {author} {\bibinfo {author} {\bibfnamefont {J.}~\bibnamefont {Gambetta}}\ and\ \bibinfo {author} {\bibfnamefont {H.~M.}\ \bibnamefont {Wiseman}},\ }\bibfield  {title} {\bibinfo {title} {Interpretation of non-{{Markovian}} stochastic {{Schr\"odinger}} equations as a hidden-variable theory},\ }\href {https://doi.org/10.1103/PhysRevA.68.062104} {\bibfield  {journal} {\bibinfo  {journal} {Physical Review A}\ }\textbf {\bibinfo {volume} {68}},\ \bibinfo {pages} {062104} (\bibinfo {year} {2003})}\BibitemShut {NoStop}%
\bibitem [{\citenamefont {Breuer}\ and\ \citenamefont {Petruccione}(2007)}]{breuer2007}%
  \BibitemOpen
  \bibfield  {author} {\bibinfo {author} {\bibfnamefont {H.-P.}\ \bibnamefont {Breuer}}\ and\ \bibinfo {author} {\bibfnamefont {F.}~\bibnamefont {Petruccione}},\ }\href {https://doi.org/10.1093/acprof:oso/9780199213900.001.0001} {\emph {\bibinfo {title} {The {{Theory}} of {{Open Quantum Systems}}}}}\ (\bibinfo  {publisher} {{Oxford University Press}},\ \bibinfo {address} {{Oxford}},\ \bibinfo {year} {2007})\BibitemShut {NoStop}%
\bibitem [{\citenamefont {Matos}\ \emph {et~al.}(2022)\citenamefont {Matos}, \citenamefont {Kantorovich},\ and\ \citenamefont {Ford}}]{matos2022}%
  \BibitemOpen
  \bibfield  {author} {\bibinfo {author} {\bibfnamefont {D.}~\bibnamefont {Matos}}, \bibinfo {author} {\bibfnamefont {L.}~\bibnamefont {Kantorovich}},\ and\ \bibinfo {author} {\bibfnamefont {I.~J.}\ \bibnamefont {Ford}},\ }\href {https://doi.org/10.48550/arXiv.2205.07288} {\bibinfo {title} {Stochastic entropy production for continuous measurements of an open quantum system}} (\bibinfo {year} {2022}),\ \Eprint {https://arxiv.org/abs/2205.07288} {arxiv:2205.07288 [cond-mat, physics:quant-ph]} \BibitemShut {NoStop}%
\bibitem [{\citenamefont {Clarke}(2022)}]{clarke2022}%
  \BibitemOpen
  \bibfield  {author} {\bibinfo {author} {\bibfnamefont {C.~L.}\ \bibnamefont {Clarke}},\ }\emph {\bibinfo {title} {Irreversibility {{Measures}} in a {{Quantum Setting}}}},\ \href@noop {} {\bibinfo {type} {Doctoral}},\ \bibinfo  {school} {UCL (University College London)} (\bibinfo {year} {2022})\BibitemShut {NoStop}%
\bibitem [{\citenamefont {Gross}\ \emph {et~al.}(2018)\citenamefont {Gross}, \citenamefont {Caves}, \citenamefont {Milburn},\ and\ \citenamefont {Combes}}]{gross2018}%
  \BibitemOpen
  \bibfield  {author} {\bibinfo {author} {\bibfnamefont {J.~A.}\ \bibnamefont {Gross}}, \bibinfo {author} {\bibfnamefont {C.~M.}\ \bibnamefont {Caves}}, \bibinfo {author} {\bibfnamefont {G.~J.}\ \bibnamefont {Milburn}},\ and\ \bibinfo {author} {\bibfnamefont {J.}~\bibnamefont {Combes}},\ }\bibfield  {title} {\bibinfo {title} {Qubit models of weak continuous measurements: Markovian conditional and open-system dynamics},\ }\href {https://doi.org/10.1088/2058-9565/aaa39f} {\bibfield  {journal} {\bibinfo  {journal} {Quantum Science and Technology}\ }\textbf {\bibinfo {volume} {3}},\ \bibinfo {pages} {024005} (\bibinfo {year} {2018})}\BibitemShut {NoStop}%
\bibitem [{\citenamefont {Helmer}\ \emph {et~al.}(2009)\citenamefont {Helmer}, \citenamefont {Mariantoni}, \citenamefont {Solano},\ and\ \citenamefont {Marquardt}}]{helmer2009a}%
  \BibitemOpen
  \bibfield  {author} {\bibinfo {author} {\bibfnamefont {F.}~\bibnamefont {Helmer}}, \bibinfo {author} {\bibfnamefont {M.}~\bibnamefont {Mariantoni}}, \bibinfo {author} {\bibfnamefont {E.}~\bibnamefont {Solano}},\ and\ \bibinfo {author} {\bibfnamefont {F.}~\bibnamefont {Marquardt}},\ }\bibfield  {title} {\bibinfo {title} {Quantum nondemolition photon detection in circuit {{QED}} and the quantum {{Zeno}} effect},\ }\href {https://doi.org/10.1103/PhysRevA.79.052115} {\bibfield  {journal} {\bibinfo  {journal} {Physical Review A}\ }\textbf {\bibinfo {volume} {79}},\ \bibinfo {pages} {052115} (\bibinfo {year} {2009})}\BibitemShut {NoStop}%
\bibitem [{\citenamefont {It{\^o}}(1944)}]{ito1944}%
  \BibitemOpen
  \bibfield  {author} {\bibinfo {author} {\bibfnamefont {K.}~\bibnamefont {It{\^o}}},\ }\bibfield  {title} {\bibinfo {title} {Stochastic integral},\ }\href {https://doi.org/10.3792/pia/1195572786} {\bibfield  {journal} {\bibinfo  {journal} {Proceedings of the Imperial Academy}\ }\textbf {\bibinfo {volume} {20}},\ \bibinfo {pages} {519} (\bibinfo {year} {1944})}\BibitemShut {NoStop}%
\bibitem [{\citenamefont {Aerts}\ and\ \citenamefont {{Sassoli de Bianchi}}(2014)}]{aerts2014}%
  \BibitemOpen
  \bibfield  {author} {\bibinfo {author} {\bibfnamefont {D.}~\bibnamefont {Aerts}}\ and\ \bibinfo {author} {\bibfnamefont {M.}~\bibnamefont {{Sassoli de Bianchi}}},\ }\bibfield  {title} {\bibinfo {title} {The extended {{Bloch}} representation of quantum mechanics and the hidden-measurement solution to the measurement problem},\ }\href {https://doi.org/10.1016/j.aop.2014.09.020} {\bibfield  {journal} {\bibinfo  {journal} {Annals of Physics}\ }\textbf {\bibinfo {volume} {351}},\ \bibinfo {pages} {975} (\bibinfo {year} {2014})}\BibitemShut {NoStop}%
\bibitem [{\citenamefont {Lukach}\ and\ \citenamefont {Smorodinskij}(1978)}]{lukach1978}%
  \BibitemOpen
  \bibfield  {author} {\bibinfo {author} {\bibfnamefont {I.}~\bibnamefont {Lukach}}\ and\ \bibinfo {author} {\bibfnamefont {{\relax Ya.A}.}~\bibnamefont {Smorodinskij}},\ }\bibfield  {title} {\bibinfo {title} {On algebra of {{Gell-Mann}}'s matrices for {{SU}}(3) group},\ }\href@noop {} {\bibfield  {journal} {\bibinfo  {journal} {Yadernaya Fizika}\ }\textbf {\bibinfo {volume} {27}},\ \bibinfo {pages} {1694} (\bibinfo {year} {1978})}\BibitemShut {NoStop}%
\bibitem [{\citenamefont {Kloeden}\ and\ \citenamefont {Platen}(1992)}]{kloeden1992}%
  \BibitemOpen
  \bibfield  {author} {\bibinfo {author} {\bibfnamefont {P.~E.}\ \bibnamefont {Kloeden}}\ and\ \bibinfo {author} {\bibfnamefont {E.}~\bibnamefont {Platen}},\ }\bibfield  {title} {\bibinfo {title} {Stochastic {{Differential Equations}}},\ }in\ \href {https://doi.org/10.1007/978-3-662-12616-5_4} {\emph {\bibinfo {booktitle} {Numerical {{Solution}} of {{Stochastic Differential Equations}}}}},\ \bibinfo {series and number} {Applications of {{Mathematics}}},\ \bibinfo {editor} {edited by\ \bibinfo {editor} {\bibfnamefont {P.~E.}\ \bibnamefont {Kloeden}}\ and\ \bibinfo {editor} {\bibfnamefont {E.}~\bibnamefont {Platen}}}\ (\bibinfo  {publisher} {{Springer}},\ \bibinfo {address} {{Berlin, Heidelberg}},\ \bibinfo {year} {1992})\ pp.\ \bibinfo {pages} {103--160}\BibitemShut {NoStop}%
\bibitem [{mov({\natexlab{a}})}]{moviespin1}%
  \BibitemOpen
  \href@noop {} {\bibinfo {title} {{{https://drive.google.com/file/d/}} {{1xcBIlHtQ-pseZNxMxSvL-PG60BxXxLNH}}/view? usp=sharing {{Movie}} of spin 1 dynamics with measurement strength 3.}} ({\natexlab{a}})\BibitemShut {NoStop}%
\bibitem [{mov({\natexlab{b}})}]{moviespin32}%
  \BibitemOpen
  \href@noop {} {\bibinfo {title} {{{https://drive.google.com/file/d/}} {{1sNncxc19eDs8ZXe5Pt66LE}}\_{{fIHOMlT}}\_v/view?usp= sharing {{Movie}} of spin 3/2 dynamics with measurement strength 1.}} ({\natexlab{b}})\BibitemShut {NoStop}%
\end{thebibliography}%

\appendix

\section{Gell-Mann matrices\label{sec:Appendix_A}}

The Gell-Mann matrices used to form the spin 1 density matrix in Eq.
(\ref{eq:rho_spin_1}) are:

\begin{align}
\lambda_{1} & =\begin{pmatrix}0 & 1 & 0\\
1 & 0 & 0\\
0 & 0 & 0
\end{pmatrix}\,\lambda_{2}=\begin{pmatrix}0 & -i & 0\\
i & 0 & 0\\
0 & 0 & 0
\end{pmatrix}\,\lambda_{3}=\begin{pmatrix}1 & 0 & 0\\
0 & -1 & 0\\
0 & 0 & 0
\end{pmatrix}\nonumber \\
\lambda_{4} & =\begin{pmatrix}0 & 0 & 1\\
0 & 0 & 0\\
1 & 0 & 0
\end{pmatrix}\,\lambda_{5}=\begin{pmatrix}0 & 0 & -i\\
0 & 0 & 0\\
i & 0 & 0
\end{pmatrix}\,\lambda_{6}=\begin{pmatrix}0 & 0 & 0\\
0 & 0 & 1\\
0 & 1 & 0
\end{pmatrix}\nonumber \\
\lambda_{7} & =\begin{pmatrix}0 & 0 & 0\\
0 & 0 & -i\\
0 & i & 0
\end{pmatrix}\,\lambda_{8}=\frac{1}{\sqrt{3}}\begin{pmatrix}1 & 0 & 0\\
0 & 1 & 0\\
0 & 0 & -2
\end{pmatrix}.\label{eq:Gell_Mann_matrices}
\end{align}

\section{Spin 1 SDEs\label{sec:Appendix_B}}

It\^{o} processes for the variables parametrising the spin 1 density
matrix are as follows:

\begin{align}
ds & =\epsilon\frac{k}{\sqrt{2}}dt-\alpha^{2}\frac{s}{2}dt-\alpha\frac{s}{3}(-3+2\sqrt{3}u+6z)dW\nonumber \\
dm & =-\epsilon\frac{2u+v}{\sqrt{2}}dt-\alpha^{2}\frac{m}{2}dt-\alpha\frac{m}{3}(-3+2\sqrt{3}u+6z)dW\nonumber \\
du & =\epsilon\frac{2m-y}{\sqrt{2}}dt+\alpha\frac{1}{\sqrt{3}}(1-2u^{2}+\sqrt{3}u(1-2z)+z)dW\nonumber \\
dv & =\epsilon\frac{m-y}{\sqrt{2}}dt-2\alpha^{2}vdt-\frac{2}{3}\alpha v(\sqrt{3}u+3z)dW\nonumber \\
dk & =\epsilon\frac{-s+x}{\sqrt{2}}dt-2\alpha^{2}kdt-\frac{2}{3}\alpha k(\sqrt{3}u+3z)dW\nonumber \\
dx & =-\epsilon\frac{k}{\sqrt{2}}dt-\frac{1}{2}\alpha^{2}xdt-\frac{1}{3}\alpha x(3+2\sqrt{3}u+6z)dW\nonumber \\
dy & =\epsilon\frac{u+v-\sqrt{3}z}{\sqrt{2}}dt-\frac{1}{2}\alpha^{2}ydt-\frac{1}{3}\alpha y(3+2\sqrt{3}u+6z)dW\nonumber \\
dz & =\epsilon\sqrt{\frac{3}{2}}ydt-\frac{1}{3}\alpha(-1+2z)(3+\sqrt{3}u+3z)dW.\label{eq:eqns_motion_spin_1}
\end{align}

\section{Spin 3/2 generators\label{sec:Appendix_C}}

The SU(2)$\otimes$SU(2) generators used to form the spin 3/2 density
matrix in Eq. (\ref{eq:rho_spin_3/2}) are as follows: $\Sigma_{1}=I_{2}\otimes\sigma_{x}$,
$\Sigma_{2}=I_{2}\otimes\sigma_{y}$, $\Sigma_{3}=I_{2}\otimes\sigma_{z}$,
$\Sigma_{4}=\sigma_{x}\otimes I_{2}$, $\Sigma_{5}=\sigma_{x}\otimes\sigma_{x}$,
$\Sigma_{6}=\sigma_{x}\otimes\sigma_{y}$, $\Sigma_{7}=\sigma_{x}\otimes\sigma_{z}$,
$\Sigma_{8}=\sigma_{y}\otimes I_{2}$, $\Sigma_{9}=\sigma_{y}\otimes\sigma_{x}$,
$\Sigma_{10}=\sigma_{y}\otimes\sigma_{y}$, $\Sigma_{11}=\sigma_{y}\otimes\sigma_{z}$,
$\Sigma_{12}=\sigma_{z}\otimes I_{2}$, $\Sigma_{13}=\sigma_{z}\otimes\sigma_{x}$,
$\Sigma_{14}=\sigma_{z}\otimes\sigma_{y}$ and $\Sigma_{15}=\sigma_{z}\otimes\sigma_{z}$,
where $I_{2}$ is a $2\times2$ identity matrix. Explicitly:

\begin{align}
 & \Sigma_{1}=\begin{pmatrix}0 & 1 & 0 & 0\\
1 & 0 & 0 & 0\\
0 & 0 & 0 & 1\\
0 & 0 & 1 & 0
\end{pmatrix}\,\Sigma_{2}=\begin{pmatrix}0 & -i & 0 & 0\\
i & 0 & 0 & 0\\
0 & 0 & 0 & -i\\
0 & 0 & i & 0
\end{pmatrix}\nonumber \\
 & \Sigma_{3}=\begin{pmatrix}1 & 0 & 0 & 0\\
0 & -1 & 0 & 0\\
0 & 0 & 1 & 0\\
0 & 0 & 0 & -1
\end{pmatrix}\,\Sigma_{4}=\begin{pmatrix}0 & 0 & 1 & 0\\
0 & 0 & 0 & 1\\
1 & 0 & 0 & 0\\
0 & 1 & 0 & 0
\end{pmatrix}\nonumber \\
 & \Sigma_{5}=\begin{pmatrix}0 & 0 & 0 & 1\\
0 & 0 & 1 & 0\\
0 & 1 & 0 & 0\\
1 & 0 & 0 & 0
\end{pmatrix}\,\Sigma_{6}=\begin{pmatrix}0 & 0 & 0 & -i\\
0 & 0 & i & 0\\
0 & -i & 0 & 0\\
i & 0 & 0 & 0
\end{pmatrix}\nonumber \\
 & \Sigma_{7}=\begin{pmatrix}0 & 0 & 1 & 0\\
0 & 0 & 0 & -1\\
1 & 0 & 0 & 0\\
0 & -1 & 0 & 0
\end{pmatrix}\,\Sigma_{8}=\begin{pmatrix}0 & 0 & -i & 0\\
0 & 0 & 0 & -i\\
i & 0 & 0 & 0\\
0 & i & 0 & 0
\end{pmatrix}\nonumber \\
 & \Sigma_{9}=\begin{pmatrix}0 & 0 & 0 & -i\\
0 & 0 & -i & 0\\
0 & i & 0 & 0\\
i & 0 & 0 & 0
\end{pmatrix}\,\Sigma_{10}=\begin{pmatrix}0 & 0 & 0 & -1\\
0 & 0 & 1 & 0\\
0 & -1 & 0 & 0\\
1 & 0 & 0 & 0
\end{pmatrix}\nonumber \\
 & \Sigma_{11}=\begin{pmatrix}0 & 0 & -i & 0\\
0 & 0 & 0 & i\\
i & 0 & 0 & 0\\
0 & -i & 0 & 0
\end{pmatrix}\,\Sigma_{12}=\begin{pmatrix}1 & 0 & 0 & 0\\
0 & 1 & 0 & 0\\
0 & 0 & -1 & 0\\
0 & 0 & 0 & -1
\end{pmatrix}\nonumber \\
 & \Sigma_{13}=\begin{pmatrix}0 & 1 & 0 & 0\\
1 & 0 & 0 & 0\\
0 & 0 & 0 & -1\\
0 & 0 & -1 & 0
\end{pmatrix}\,\Sigma_{14}=\begin{pmatrix}0 & -i & 0 & 0\\
i & 0 & 0 & 0\\
0 & 0 & 0 & i\\
0 & 0 & -i & 0
\end{pmatrix}\nonumber \\
 & \Sigma_{15}=\begin{pmatrix}1 & 0 & 0 & 0\\
0 & -1 & 0 & 0\\
0 & 0 & -1 & 0\\
0 & 0 & 0 & 1
\end{pmatrix}.\label{eq:su2_generators-1}
\end{align}

\section{Spin 3/2 SDEs\label{sec:Appendix_D}}

It\^{o} processes for the variables parametrising the spin 3/2 density
matrix are as follows:

\begin{align}
dv & =-\epsilon odt-\frac{1}{2}\alpha^{2}vdt+\alpha(2q-v(f+2p)dW\nonumber \\
de & =\epsilon(\sqrt{3}f+k)dt-\frac{1}{2}\alpha^{2}edt+\alpha(2s-e(f+2p))dW\nonumber \\
df & =-\epsilon(\sqrt{3}e+j-m)dt-\alpha(-1+f^{2}+2fp-2u)dW\nonumber \\
dg & =-\epsilon sdt-2\alpha^{2}gdt+\alpha(k-g(f+2p))dW\nonumber \\
dh & =-\frac{1}{2}\alpha^{2}(5h-4n)dt-\alpha h(f+2p)dW\nonumber \\
dj & =\epsilon(f+\sqrt{3}k-p)dt-\frac{1}{2}\alpha^{2}(5j+4m)dt-\alpha j(f+2p)dW\nonumber \\
dk & =-\epsilon(e+\sqrt{3}j)dt-2\alpha^{2}kdt+\alpha(g-k(f+2p))dW\nonumber \\
dl & =\epsilon qdt-2\alpha^{2}ldt+\alpha(o-l(f+2p))dW\nonumber \\
dm & =\epsilon(-f+p)dt-\frac{1}{2}\alpha^{2}(4j+5m)dt-\alpha m(f+2p)dW\nonumber \\
dn & =\epsilon\sqrt{3}odt+\alpha^{2}(2h-\frac{5n}{2})dt-\alpha n(f+2p)dW\nonumber \\
do & =\epsilon(-\sqrt{3}n+v)dt-2\alpha^{2}odt+\alpha(l-o(f+2p))dW\nonumber \\
dp & =\epsilon(j-m)dt+\alpha(2-p(f+2p)+u)dW\nonumber \\
dq & =-\epsilon ldt-\frac{1}{2}\alpha^{2}qdt-\alpha(fq+2pq-2v)dW\nonumber \\
ds & =\epsilon(g+\sqrt{3}u)dt-\frac{1}{2}\alpha^{2}sdt+\alpha(2e-s(f+2p))dW\nonumber \\
du & =-\epsilon\sqrt{3}sdt+\alpha(2f+p-u(f+2p))dW.\label{SDEs for spin 3/2}
\end{align}

\end{document}